\documentclass[12pt]{article}

\usepackage{natbib,amssymb,amsfonts,amsbsy,amsmath,hyperref,geometry}

\usepackage{colortbl}

\newcommand{\red}{\textcolor{red}}
\newcommand{\double}{\renewcommand{\baselinestretch}{1.25}\small\normalsize}

\def\references{\bibliography{JLRef-AnnReview-2-05-15C,classification141122s,15-3-1}}

\DeclareMathOperator*{\argmin}{arg\,min}
\DeclareMathOperator*{\argmax}{arg\,max}
\newcommand{\la}{\label}

\newcommand{\be}{\begin{eqnarray}}
\newcommand{\ee}{\end{eqnarray}}
\newcommand{\bea}{\begin{eqnarray*}}
\newcommand{\eea}{\end{eqnarray*}}
\newcommand{\bi}{\begin{itemize}}
\newcommand{\ei}{\end{itemize}}
\newcommand{\ben}{\begin{enumerate}}
\newcommand{\een}{\end{enumerate}}
\newcommand{\bay}{\begin{array}}
\newcommand{\eay}{\end{array}}
\newcommand{\bsl}{\begin{slide}}
\newcommand{\esl}{\end{slide}}
\newcommand{\bcen}{\begin{center}}
\newcommand{\ecen}{\end{center}}
\newcommand{\mb}{\mathbf}
\newcommand{\noi}{\noindent}
\newcommand{\var}{\rm{var}}
\newcommand{\ed}{\end{document}}
\newcommand{\bs}{\boldsymbol}

\newcommand{\bA}{{\bf A}}

\newcommand{\E}{\mbox{{\rm E}}}

\newcommand{\R}{{\mathbb{R}}}

\def\bco{\iffalse}

\def\cov{{\rm cov}}

\def\var{{\rm var}}

\def\corr{{\rm corr}}

\def\vs{\vspace{.125cm}}

\newcommand{\beq}{\begin{equation*}}
\newcommand{\eeq}{\end{equation*}}
\newcommand{\beqn}{\begin{equation}}
\newcommand{\eeqn}{\end{equation}}
\def\R{\mathbb{R}}
\def\L2{_{L^2}}

\newcommand{\mf}{\mathbf}
\def\tij1{t_{i,j-1}}

\def\mani{\mathcal{M}}

\def\umani{\bs{\mu}}

\def\ejmani{\mf{e}_j^\mani}

\def\Xmu{\mu^\mani}

\def\slj{(\lambda_j^\mani)^{\frac{1}{2}}}
\def\ci{\cite}
\def\cp{\citep}
\def\no{\noindent}

\def\bc{\begin{center}}
\def\ec{\end{center}}
\def\Xc{X(t)-\mu(t)}
\def\Xd{X^{(1)}(t)-\mu^{(1)}(t)}

\begin{document}
 \thispagestyle{empty}

\double 
\bc
{\Large \bf Review of Functional Data Analysis}
\vspace{0.15in}\\
{\large   Jane-Ling Wang,$^1$ Jeng-Min Chiou,$^2$ and Hans-Georg M{\"u}ller$^1$}\vspace{0.15in}\\
$^1$Department of Statistics, University of California, Davis, USA, 95616\\
$^2$Institute of Statistical Science, Academia Sinica,Tapei, Taiwan, R.O.C.
\ec

 \bc{\bf \sf ABSTRACT} \ec \vspace{.5in} \no 

With the advance of modern technology, more and more data are being recorded continuously during a time interval or intermittently at several discrete time points. They are both examples of ``functional data'', which have become a prevailing type of data.   Functional Data Analysis (FDA) encompasses the statistical methodology for such data. Broadly interpreted, FDA 
deals with the analysis and theory of data that are in the form of functions. This paper provides an overview of  FDA, starting with simple statistical notions such as mean and covariance functions, then covering some core techniques, the most popular of which is Functional Principal Component Analysis (FPCA). FPCA is an important dimension reduction tool and in sparse data situations can be  used to impute functional data that are sparsely observed.  Other dimension reduction approaches are also discussed. In addition, we review another core technique, functional linear regression, as well as clustering and classification of functional data. Beyond linear and single or multiple index methods we touch upon a few nonlinear approaches that are promising for certain applications. They include additive and other nonlinear functional regression models, such as time warping, manifold learning, and dynamic modeling with empirical differential equations. The paper concludes with a brief discussion of future directions.

\vs\vs

\no {KEY WORDS:\quad Functional principal component analysis, functional correlation, functional linear regression, functional additive model, clustering and classification, time warping

\newpage

\section{Introduction}
Functional data analysis (FDA) deals with the analysis and theory of data that are in the form of functions, images and shapes, or more general objects.  The atom of functional  data is a function, where for each subject in a random sample one or several functions are recorded.   While the  term ``functional data analysis"  was coined by \citet{Rams82} and \citet{RamsD91}, the history of this area is much older and dates back to \citet{gren:50} and \citet{Rao58}.  Functional data are intrinsically infinite dimensional.  The high intrinsic dimensionality of these data poses challenges both for theory and computation, where these challenges vary with how the functional data were sampled. On the other hand, the high or infinite dimensional structure of the data is a rich source of information, which brings many opportunities.  

First generation functional data typically consist of a random sample of independent real-valued functions, $X_1(t), \ldots, X_n(t)$,  on a compact interval $I=[0,T]$ on the real line. Such data have also been  termed curve data  \citep{GassMK84, RiceS91, gass:95}.  These real-valued functions  can be viewed as the realizations of  a one-dimensional stochastic process, often assumed to be in a Hilbert space, such as $L^2(I).$   Here a stochastic process $X(t)$ is said to be an $L^2$ process if  and only if it satisfies $E(\int_I X^2(t) dt ) < \infty$.  While it is possible to model functional data with  parametric approaches, usually mixed effects nonlinear models, the massive information contained in the infinite dimensional data and the need for a large degree of flexibility, combined with a  natural ordering (in time) within a curve datum facilitate non- and semi-parametric approaches, which are the prevailing methods in the literature as well as the focus of this paper.  Smoothness of the individual function (or stochastic process), such as existence of continuous second derivatives,  is often imposed for regularization, which is especially useful if nonparametric smoothing techniques are  employed, as is prevalent in functional data analysis 


In this paper, we focus on first generation functional data with brief a discussion of next generation functional data in Section 6.   Here next generation functional data refers to functional data that are part of complex data objects, and possibly are multivariate, correlated, or involve  images or  shapes. Examples of next generation functional data include brain and neuroimaging data. 
A separate entry on functional data approaches for neuroimaging data is available at the same issue of the Annual Reviews (Link to John Aston's contribution). 
For a brief discussion of next generation functional data, see page 23 of a report (http://www.worldofstatistics.org/wos/pdfs/Statistics\&Science-TheLondonWorkshopReport.pdf) of the London workshop on the Future of Statistical Sciences held in November 2013.  




Although scientific interest is in the underlying stochastic process and its properties,  in reality this process is often latent and cannot be observed directly, as data can only be collected discretely over time, either on a fixed or random time grid.   The time grid can be dense, sparse, or neither; and may vary from subject to subject.  Originally, functional data were regarded as  samples of fully observed trajectories. A slightly more general assumption is that functional data are recorded on the same dense time grid $t_1, \ldots, t_p$ for all $n$ subjects.  If the recording is done by an instrument, such as EEG or fMRI machine, the time grid is usually equally spaced, that is $t_{i+1}-t_i=t_{j+1}-t_j$ for all $i$ and $j$. In asymptotic analysis,  the spacing $t_{j+1}-t_j$ is assumed to approach zero as $n$ tends to infinity, hence $p=p_n$ is a sequence that tends to infinity. 
 On one hand, large $p$ leads to  a high-dimensional problem, 
 but also
 means more data  so should be a blessing rather than a curse.  This blessing is realized by imposing a smoothness assumption on the $L^2$ processes, so that information from measurements at neighboring time points can be pooled to overcome the curse of dimensionality. Thus, smoothing serves as a tool for regularization.

While there is no formal definition of ``dense'' functional data, the convention has been that $p_n$ has to converge to infinity fast enough to allow the corresponding estimate for the mean function $\mu(t)=EX(t)$, where $X$ is the underlying process,  to attain the parametric $\sqrt{n}$ convergence rate  for standard metrics,  such as the $L^2$ norm. Sparse functional data arise in longitudinal studies where subjects are measured at different time points and the number of measurements $n_i$ for subject $i$  may be bounded away from infinity, i.e., $\sup_{1 \le i \le n} n_i <C<\infty$ for some constant $C.$ 
A rigorous definition of the types of functional data based on their sampling plans is still lacking, see  \citep{ZhanW14} for a possible approach,  with further details in  Section 2 below.  

In reality, the observed data often are contaminated by random noise, referred to as measurement errors, which are often assumed to be independent across and within subjects.  Measurement errors can be viewed as random fluctuations around a smooth trajectory, or as actual errors in the measurements.  A strength of FDA  is that it can accommodate measurement errors easily because  for each subject one observes repeated measurements.  An interesting, but perhaps not surprising, phenomenon in FDA is that the methodology and theory, such as convergence rates, varies with the sampling plan of the time grid, i,e,,  the measurement schedule.   Intriguingly, sparse and irregularly sampled functional data, that we synonymously  refer to as longitudinal data, typically require more effort in theory and methodology as compared to densely sampled functional data.  Functional data that are, or assumed to be, observed continuously without errors are the easiest type to handle as theory for stochastic processes, such as functional laws of large numbers and functional central limit theorems, are readily applicable.
A comparison of the various approaches will be presented in Section 2, with discussion of a unified approach for various sampling plans. 

One challenge in functional data analysis is the inverse nature of functional regression and most functional correlation measures. This is triggered by the compactness of the covariance operator, which leads to unbounded inverse operators.  This challenge will be discussed further in Section 3, where  extensions of classical linear and generalized linear models to functional linear and generalized functional linear models will be reviewed. Since functional data are intrinsically infinite dimensional, dimension reduction is key for data modeling and analysis.  The principal component approach will be explored in Section 2 while several approaches for dimension reduction  regression 
will be discussed in Section 3.

Clustering and classification of functional data are useful and important tools in FDA with wide ranging applications. Methods include extensions of classical $k$-means and hierarchical clustering, Bayesian and model-based approaches to clustering, as well as functional regression based and functional discriminant analysis approaches to classification.  These topics will be explored in  Section~\ref{sec:cluster.class}.   

The classical methods for functional data analysis have been predominantly linear, such as functional principal components or the functional linear model. As more and more functional data are being generated, it has emerged that many such data have inherent nonlinear features that make linear methods less effective. Sections 5 reviews some nonlinear approaches to FDA, including time warping, non-linear manifold modeling, and nonlinear differential equations to model the empirical dynamics inherent in functional data. 

A well-known and well-studied nonlinear effect is time warping, where in addition to the common amplitude variation one also considers time variation. This creates a basic non-identifiability problem. Section 5.1 will provide a discussion of these foundational issues. 
 A more general approach to model nonlinearity in functional data that extends beyond time warping and includes many other nonlinear features that may be present in longitudinal data is to assume that the functional data lie on a nonlinear (Hilbert) manifold. The starting point for such models is the choice of a suitable distance and ISOMAP  \citep{Tene:00} or related methods can then be employed to uncover the manifold structure and define functional manifold means and components. 
These approaches will be described in Section 5.2.
Modeling of time-dynamic systems with differential equations that are learned from many realizations of the trajectories of the underlying stochastic process and the learning of nonlinear empirical dynamics such as dynamic regression to the mean or dynamic explosivity is briefly reviewed in Section 5.3. 

Section 6 concludes this review  with a brief outlook on the future of functional data analysis, where the emphasis shifts to next generation functional data.

Research tools that are useful for handing functional data include 
various smoothing methods, notably kernel, local least squares and spline smoothing for which various excellent reference books exist \citep{WandJ95, FanG96, Euba99, Boor01} and knowledge on functional analysis \citep{Conw94, RiesN90}. 
Several software packages are publicly available to analyze functional data, including software at the Functional Data Analysis website of James Ramsay (http://www.psych.mcgill.ca/misc/fda/),  the fda package on the crane project of R \newline (http://cran.r-project.org/web/packages/fda/fda.pdf),  \newline the Matlab package PACE on the website of the Statistics Department of the University of California, Davis (http://www.stat.ucdavis.edu/PACE/), and the R  package refund on functional regression
(http://cran.r-project.org/web/packages/refund/index.html).   


This review is  based on a subjective selection of topics in FDA that the authors have worked on or find of particular interest. We do not attempt to provide an objective or comprehensive review of this fast moving field and apologize in advance for any omissions of relevant work.  By now there are many alternative approaches to handle functional data.  Interested readers can explore the various aspects of this field   through several monographs \citep{Bosq00, 
rams:05, ferr:06, WuZ06, RamsHG09, horv:12, HsinE15} and review articles \citep{Rice04, ZhaoMW04, Mull05, mull:08:7,  ferr:06}.   Several special journal  issues were devoted to FDA including a 2004 issue of {\it Statistica Sinica} (issue 3), a 2007 issue in {\it Computational Statistics and Data Analysis} (issue 3), and a  2010  issue in {\it Journal of Multivariate analysis} (issue 2).

\bco 

{\color{red} Maybe remove the next paragraph or move to later?}

In addition to consider functional data as $L^2$-processes, some restrict the functional data to reside in a reproducing kernel Hilbert space (RKHS), which  is smaller than the $L^2$ space. This restriction, similar to choosing a smaller parameter space,  has the technical advantage that  it can overcome some of the  challenges of inverse problem that an $L^2$ space faces and may lead to a faster convergence rate \citep{YuanC10}. {\color{red}  ? Put this later in sec 2.}  However, it must be bore in mind that this is at the cost of a more restrictive model. For theoretical foundations of FDA and the RKHS approach, the new book by \citet{HsinE15}  is an excellent resource.

\fi

\section{Mean and Covariance Function, and Functional Principal Component Analysis}
In this section, we focus on first generation functional data  that are i.i.d. realizations of a stochastic process $X$ that is in $L^2$  and defined on the interval $I$ with mean function $\mu (t)=E(X(t))$ and covariance function  $\Sigma (s, t)= \mbox{cov} (X(s), X(t))$. The functional framework can also be extended to $L^2$ processes with multivariate arguments.  The realization of the process for the $i$th subject is $X_i=X_i(\cdot)$, and the sample consists of $n$ subjects. For generality, we allow the sampling schedules  to vary  across subjects and denote the sampling schedule for subject $i$ as  $t_{i1}, \ldots, t_{in_i}$ and the corresponding 
observations  as ${\bf X}_i = (X_{i1}, \ldots, X_{in_i})$, where $X_{ij}=X_i (t_{ij})$.  In addition, we allow the measurement of  $X_{ij}$  to be contaminated by a random noise $e_{ij}$ with $E(e_{ij})=0$ and  $\mbox{var}(e_{ij})= \sigma^2_{ij}$, so the actual observed value is $Y_{ij}= X_{ij}+e_{ij}$,  where $e_{ij}$ are independent across $i$  and $j$  and often termed ``measurement errors''.  

It is often assumed that  the errors are homoscedastic with $\sigma^2_{ij}=\sigma^2$, but this is is not strictly necessary,  as long as $\sigma^2_{ij} = \mbox{ var} (e(t_{ij})) $ can be regarded as the discretization of a smooth variance function $\sigma^2(t)$.   
We observe that measurement errors are realized only  at those time points $t_{ij}$ where measurements are being taken.     Hence these errors  do not form a stochastic process $e(t)$ but rather should be treated as discretized data $e_{ij}$.   However, in order to estimate the variance $\sigma^2_{ij}$ of $e_{ij}$ it is often convenient to assume that there is a latent smooth function $\sigma(t)$  such that $\sigma_{ij} =\sigma^2(t_{ij})$.   \\ 


\noindent {\bf Estimation of Mean and Covariance Functions.} \  When subjects are sampled at the same time schedule, i.e., $t_{ij}=t_j$  and $n_i=m$ for all $i$, the observed data are $m$-dimensional multivariate data, so  the mean and covariance  can be estimated empirically at the measurement times  by the sample mean and sample covariance, $\hat{\mu}(t_j) = \frac{1}{n}  \sum_{i=1}^n Y_{ij}$, and $\hat{\Sigma}(t_k, t_l)=  \frac{1}{n} \sum_{i=1}^n  (Y_{ik} - \hat{\mu}(t_{ik})) (Y_{il} -\hat{\mu}(t_{il}))$, for $k\neq l$.  Missing data (missing completely at random) can be handled easily by adjusting the available sample size at each time point $t_j$  for the mean estimate or by adjusting the sample sizes of available pairs at  $(t_k, t_l)$ for the covariance estimate. 
An estimate of the mean and covariance functions on the entire interval $I$ can then be obtained by  smooth interpolation of the corresponding sample estimates or by mildly smoothing over the grid points. Such a smoothing step enhances the global estimate of the mean and auto-covariance functions and consistency can be attained only if $m=m_n$ grows with the sample size and approaches infinity. 
Once we have a smoothed estimate $\hat{\Sigma}$ of  the covariance function $\Sigma$, the variance of the measurement error at time $t_j$ can be estimated as $\hat{\sigma}^2(t_j) =  \frac{1}{n} \sum_{i=1}^n  (Y_{ij} - \hat{\mu}(t_{j}))^2 - \hat{\Sigma}(t_{j}, t_{j})$, because  var$(Y(t))=$ var$(X(t)) + \sigma^2 (t)$.

When the sampling schedule of subjects differs, the above sample estimates cannot be obtained.  However, one can borrow information from neighboring data and across all subjects to estimate  the mean  function, provided the sampling design combining all subjects, i.e. $ \{t_{ij}: \, 1 \le  i  \le  n,\, 1 \le j \le n_i\}$, is a dense subset of the interval $I$.  Then  a nonparametric smoother, such as a local polynomial estimate \citep{FanG96},  can be applied to the scatter plot $\{(Y_{ij}, t_{ij}):  i=1, \ldots, n,  \  \mbox{and} \ j=1, \ldots, n_i\}$ to smooth $Y_{ij}$ against $t_{ij}$ across time and will yield consistent estimates of $\mu(t)$ for all $t$. 
 Likewise, the covariance  can be estimated on $I \times I$ by a two-dimensional scatter plot smoother $\{(u_{ikl}, t_{ik}, t_{il}): i=1, \ldots, n;  \ k,  l=1,\ldots, n_i, k \neq l \}$ to smooth $u_{ikl}$ against $(t_{ik}, t_{il})$ across the two dimensional product time intervals, where $u_{ikl}= (Y_{ik}- \hat{\mu}(t_{ik}) )  (Y_{il}- \hat{\mu}(t_{il}) )  $ are the raw covariances.
 We note that  the diagonal raw covariances where $k=l$ are removed from the 2D scatter plot prior to the smoothing step because these include an additional term that is due to the variance of the  measurement errors in the observed $Y_{ij}$.  Indeed, once an estimate $\hat{\Sigma}$ for $\Sigma$ is obtained, the variance $\sigma^2(t)$  of the measurement errors can be obtained by smoothing $Y_{ij}- \hat{\mu}(t_{ij})^2- \hat{\Sigma}(t_{ij})$ against $t_{ij}$ across time. A better estimate for $\sigma$ under the homoscedasticity assumption is discussed in \citet{YaoMW05a}.  
 
 
 The above smoothing approach is based on a scatter plot smoother which assigns equal weights to each observation, therefore subjects with a larger number of repeated observations receive more total weight, and  hence contribute more toward the estimates of the  mean and covariance functions.  An alternative approach employed in \citet{LiH10}  is to assign equal weights to each subject.  Both approaches are sensible. A  question is which one would be preferred for a particular design and whether there is a unified way to deal with these two methods and their theory.   These issues were recently explored in a manuscript \citep{ZhanW14}, employing a general weight function and providing  a comprehensive analysis of the asymptotic properties on a unified platform for three types of asymptotics, $L^2$ and $L^{\infty}$ (uniform) convergence as well as asymptotic normality of the general weighted estimates. Functional data sampling designs are further  partitioned into three categories, non-dense (designs where one cannot attain the $\sqrt{n}$ rate), dense (where one can attain the $\sqrt{n}$ rate but with a non-neglible asymptotic bias), and ultra-dense (where one can attain the $\sqrt{n}$ rate without asymptotic bias).  Sparse sampling scenarios where  $n_i$ is uniformly bounded by a finite constant are a  special case of non-dense data and lead to  the slowest convergence rates. These designs are also referred to as longitudinal designs. The differences in the convergence rates also have ramifications  for the construction of simultaneous confidence bands.  For ultra dense or some dense functional data, the weighing scheme that assigns equal weights to subjects is generally more efficient than the scheme that assigns equal weight per observation but the situation is reversed for many other sampling plans, including sparse functional data. 
 
\medskip 
 
 \noi \textbf{Hypothesis Testing and Simultaneous Confidence Bands  for Mean and Covariance Functions.}
  Hypothesis testing for the comparison of mean functions $\mu$ is of obvious interest.   
 \citet{FanL98}   proposed a two-sample test and ANOVA test  for the mean functions, with further work  by \citet{CuevFF04} and  \citet{Zhan13}. Other two sample tests were studied for distributions of functional data  \cp{HallVK07} and for the covariance functions \cp{pana:10,BoenRS11}.   
  
    
   
 Another inference problem that has been  explored is the construction of   
 simultaneous confidence bands for dense \cp{Degr08, Degr11, WangY09,CaoYT12} and sparse  \cp{MaYC12} functional data. 
However,  the problem has not been completely resolved for functional data due to two main obstacles, the infinite dimensionality of the data and the nonparametric nature of the target function. For the mean function $\mu$, an interesting ``phase transition" phenomenon emerges: For ultra-dense data the estimated mean process $\sqrt{n} (\hat{\mu}(t)-\mu(t))$  converges to a mean zero Gaussian process $W(t)$,  for $ t \in I,$ so standard continuous mapping leads to a construction of a  simultaneous confidence band based on the distribution of $\sup_t W(t)$.    When the functional data are dense but not ultra dense, the process  $\sqrt{n} (\hat{\mu}(t)-\mu(t))$  can still converge to a Gaussian process $W(t)$ with a proper choice of smoothing parameter but $W$ is no longer centered at zero due to the existence of asymptotic bias as discussed in Section 1. 

 This resembles the classical situation of estimating a regression function, say  $m(t)$,  based on independent scalar response data, where there is a trade off between the bias and variance so optimally smoothed  estimates of the regression function will have an asymptotic bias. The conventional approach to  construct a pointwise confidence interval  is based on the distribution of $r_n (\hat{m}(t) - E(\hat{m}(t)))$, 
 where $\hat{m}(t)$ is an estimate of $m(t)$ at the optimal rate $r_n$. 
 This means that the asymptotic confidence interval derived from it is targeting $E(\hat{m}(t))$ rather than the true target $m(t)$ and therefore is not really viable for inference. 
 
 \bco 
 
    A similar situation arises in the functional context when $m$ is replaced by the mean function $\mu$ and the responses are  functions.  The pointwise confidence interval is based on the asymptotic distribution of $\sqrt{n} (\hat{\mu}(t) - E(\hat{\mu}(t)))$ rather than $\sqrt{n} (\hat{\mu}(t)-\mu(t))$. Other than this challenge triggered by the bias-variance trade off of a nonparametric smoothing procedure, simultaneous confidence band can be constructed based on $sup_t W^*(t)$, where $W^*$ is the limiting process of $\sqrt{n} (\hat{\mu}(t) - E(\hat{\mu}(t)))$, for $t \in I$.  \citet{BuneIW11} provided a partial solution to this problem when the random function $X$  is a Gaussian process.  
 
{\color{red} Hans- pls check the discussion on the centering issue.}

  Simultaneous confidence bands for the mean function $\mu$ based on sparsely observed  functional data requires a different strategy,
as  neither the mean process,  $\sqrt{n} (\hat{\mu}(t)-\mu(t))$  nor  $\sqrt{n} (\hat{\mu}(t)- E(\hat{\mu}(t)))$ are tight. The approach of \citet{BickR73} for nonparamtric density function estimation is still applicable. 
so the construction of the simultaneous confidence band 
 follows the line of work by \citet{BickR73} for nonparamtric density function estimation, which is also shared by   nonparametric regression function estimation.  

\fi 
 
 In summary, the construction of simultaneous confidence band for functional data requires  different methods for ultra-dense, dense, and sparse functional data, where in the latter case one does not have tightness and the rescaling approach of \citet{BickR73} may be applied. The divide between the various sampling designs is perhaps not unexpected since ultra dense functional data falls along the paradigm of parametric inference where the $\sqrt{n}$ rate of convergence is attained with no asymptotic bias, while dense functional data attains the parametric rate of $\sqrt{n}$ convergence albeit with an asymptotic bias, which leads to challenges even in the construction of  pointwise confidence intervals.  Unless the bias is  estimated separately, removed from the limiting distribution, and proper asymptotic theory is established, which usually requires regularity conditions for which the estimators are not efficient, the resulting confidence intervals need to be taken with a grain of salt. This issue is specific to the bias-variance trade off that   is inherited from  nonparametric smoothing. Sparse functional data follow a very  different paradigm as they allow no more than  nonparametric convergence rates, which are slower than  $\sqrt{n}$, and the rates depend on the design of the measurement schedule and properties of mean and covariance function as well as the smoother \cp{ZhanW14}. The phenomenon of  nonparametric versus parametric convergence rates  as designs get more regular and denser characterize a sharp ``phase transition'' \cp{mull:06:7,cai:11}.

  


  
  \medskip
 \noindent {\bf Functional Principal Component Analysis (FPCA).} \  Principal component analysis \citep{Joll02} is a key dimension reduction tool for multivariate data that has been extended to functional data and termed functional principal component analysis (FPCA).  Although the basic ideas were conceived in \citet{gren:50, Karh46, Loev46} and \citet{Rao58}, a more comprehensive  framework for statistical inference  for FPCA was first developed in a joint Ph.D. thesis of Dauxois and Pousse (1976) at the University of Toulouse \citep{DauxPR82}.    Since then, this approach has taken off to
 become the most prevalent tool in FDA. This is partly because  FPCA facilitates the conversion of inherently infinite-dimensional functional data to a finite-dimensional vector of random scores. Under mild assumptions, the  underlying stochastic process can be expressed as a countable sequence of uncorrelated random variables, the functional principal  components (FPCs) or scores, which are then truncated to a finite vector. Then the tools of multivariate data analysis can be readily applied to the resulting random vector of scores, thus accomplishing the goal of dimension reduction.

 Specifically, the dimension reduction is achieved through an expansion of the underlying but often not fully observed random trajectories $X_i(t)$ in a functional basis that consists of the eigenfunctions of the auto-covariance operator of the process $X$.  With a slight abuse of notation we define the covariance operator as 
 $\Sigma(g)= \int_I \Sigma (s, t) g(s) ds, $ for any function $g \in L^2$, using the same notation for the covariance operator and covariance  function.  Because of the integral form, the covariance operator is a trace class and hence  compact Hilbert-Schmidt operator  \citep{Conw94}.  It also has real-valued nonnegative eigenvalues $\lambda_j$, because it is symmetric and non-negative definite. Under mild assumptions, Mercer's theorem implies that the spectral decomposition of $\Sigma$ leads to $\Sigma(s,t)=  \sum_{k=1}^\infty  \lambda_k \phi_k(s) \phi_k(t),$ with uniform convergence, where  $\lambda_k$ are the eigenvalues (in descending order) of the covariance operator and $\phi_k$  the corresponding orthogonal eigenfunctions.   Karhunen  and Lo{\`e}ve   \citep{Karh46, Loev46}  independently discovered the  FPCA expansion
 \be X_i(t)= \mu(t) + \sum_{k=1}^{\infty} A_{ik} \phi_k(t), \label{KL}
 \ee
 where $A_{ik}= \int_I (X_i(t)-\mu(t)) \phi_k (t) dt$ are the functional principal components (FPCs) of $X_i$. The  $A_{ik}$  are independent  across $i$ for a sample of independent trajectories and are uncorrelated across $k$ with $E(A_{ik})=0\,  \var(A_{ik})=\lambda_k.$  The convergence of the sum in (\ref{KL}) is with respect to the $L^2$ norm.  
 Expansion (\ref{KL}) facilitates dimension reduction as the first $K$ terms for large enough $K$  provide a good approximation to the infinite sum and therefore for  $X_i$, so that the information contained in $X_i$ is essentially contained in the  $K$-dimensional vector $\bA_i = (A_{i1}, \ldots, A_{iK})$ and one works with the approximated processes 
  \be X_{iK}(t)= \mu(t) + \sum_{k=1}^{K} A_{ik} \phi_k(t), \label{KLK}
 \ee 
 
  Analogous dimension reduction can be achieved by expanding the functional data into other  function bases, such as spline, Fourier, or wavelet bases.  What distinguishes FPCA is that for a given number of $K$ components the expansion that uses these components explains most of the variation in $X$ in the $L^2$ sense.  When choosing $K$ in an estimation setting, there is a trade off between bias (which gets smaller as $K$ increases due to the smaller approximation error) and variance (which increases with $K$ as more components must be estimated, adding random error). So a model selection procedure is needed, where  typically $K=K_n$ is 
 considered to be a function of sample size $n$ and $K_n$ must tend to infinity to obtain consistency of the representation.   Because of this, the theory for FPCA is quite different from standard multivariate analysis theory.  
 
     The estimation of the eigencomponents (eigenfunctions and eigenvalues) of FPCA is  straightforward once the mean and covariance of the functional data have been obtained. To obtain the spectral decomposition of the covariance operator, which yields the eigencomponents, one simply approximates the estimated auto-covariance surface $\cov(X(s), X(t))$  on a  finite grid, thus reducing  the problem  to the corresponding matrix spectral decomposition.  The convergence of the estimated eigencomponents is obtained by combining results on the convergence of the covariance estimates that are achieved under regularity conditions with  perturbation theory (see Chapter VIII of \citet{Kato80}). 
     
 For situations where  the covariance surface cannot be estimated at the $\sqrt{n}$ rate,  the convergence of estimates is typically influenced by the smoothing method that is employed. Consider the sparse case, where the convergence rate of the covariance surface corresponds to the optimal rate at which a smooth  two-dimensional surface can be estimated. Intuition suggests that the eigenfunction, which is a one-dimensional function, should be estimable at the one-dimensional optimal rate for smoothing methods. An affirmative answer is provided in \citet{mull:06:7}, where eigenfunction estimates were shown to attain the better (one-dimensional) rate of convergence, if one is undersmoothing the covariance surface estimate.  
 This phenomenon resembles a scenario encountered in semiparametric inference, where a $\sqrt{n}$ rate is attainable for the parametric component if one undersmooths the nonpararmetric component before estimating the parametric component.  This undersmoothing can be avoided so that the same smoothing parameter can be employed for both the parametric and nonparametric component if a profile approach is employed to estimate the parametric component.  An interesting and still open question is how to construct such a profile approach so  that the eigenfunction is the direct target of the estimation procedure, bypassing the estimation of the covariance function.  
  
    Another open question is the choice of the number of components $K$ needed for the approximation (\ref{KLK}) of the full  Karhunen-Lo{\'e}ve  expansion (\ref{KL}) for various applications of FPCA. 
 There are several ad hoc procedures that are  routinely applied in multivariate PCA, such as  the scree plot or the fraction of variance explained by the first few PC components, which can be directly extended to the functional setting. Other approaches are  pseudo-versions of AIC (Akaike information criterion) and BIC (Bayesian information criterion) \citep{YaoMW05a}, where the latter selects fewer components. Cross-validation with one-curve-leave-out has also been investitgated  \citep{RiceS91}, but tends to overfit  functional data by selecting too large  $K$ in (\ref{KLK}).   
 Additional work for model selection is needed, where an initial analysis is due to 
     \citet{HallV06}, who  studied a special case where the functional data are of finite dimension, i.e. the total number of components in (\ref{KL}) is  a finite integer rather  than $\infty$.  In addition to the order $K$ one also needs to choose various tuning parameters for the smoothing steps and their optimal selection in the context of FDA remains a challenge due to the auto-correlations  within a subject. 
     
     FPCA for fully observed functional data was studied in \citet{DauxPR82}, \citet{BessR86, Silv96}, \citet{Bosq00,  BoenF00, HallH06}.  FPCA for  densely observed functional data was explored in  \citet{cast:86, RiceS91, PezzS93}  and \citet{Card00}.   For the much more difficult but commonly encountered situation of sparse functional data, the FPCA approach was investigated in \citet{ShiWT96, StanL98, JameHS00,  RiceW01, YaoMW05a, YaoL06};  and  \citet{PaulP09}.    
 The FPCA approach has also been extended to incorporate covariates \cp{ChioMW03a,Card07,  mull:09:3} for vector covariates  and dense functional data, 
 and also  for sparse functional data with vector or functional covariates 
  \cp{JianW10, JianW11} . 
 
 The aforementioned approaches of FPCA are not robust against outliers because  principal component analysis involves second order moments.  Outliers for functional data have many different facets due to the high dimensionality of these data.  They can appear as outlying measurements at  a single or several time points, or as an outlying  shape of an entire  function. 
 Current approaches to deal with outliers and contamination and more generally visual exploration of functional data include  exploratory box plots \citep{HyndS10, SunG11} and  
 robust versions of FPCA  \cp{
 CramDL08, Gerv08, BaliBT11, KrauP12, BoenS14}.    More research on outlier detection and robust FDA approaches are needed. 

\medskip

\noi \textbf{Applications of FPCA.}   \ 
The FPCA approach motivates the concept of modes of variation for functional data \cp{jone:92}, a most useful tool to visualize and describe  the variation in the functional data that is contributed by each eigenfunction. The  $k-$th mode of variation is the set of functions $$\mu(t) \pm \alpha \sqrt{\lambda_k} \phi_k(t),\, \,
t \in I, \,\, \alpha \in [-A,A],$$ that are viewed simultaneously over the range of $\alpha$, usually for  $A=2$, substituting estimates for the unknown quantities. 
Often the eigencomponents and associated modes of variation have compelling and sometimes striking interpretations, such as for the evolution of functional traits \cp{kirk:89} and in many other applications \cp{knei:01,  RamsS02}. 
  FPCA also facilitates functional principal component regression by mapping the function to its first few principal components, then employing regression models with  vector predictors. Since FPCA is an essential dimension reduction tool, it is also useful for  classification and clustering of functional data (see Sections 3). 

Last  but not least, FPCA facilitates the construction of  parametric models that will be more parsimonious. For instance, if the first two principal components   explain over 90\% of the variation of the data, then one can approximate the original functional data with only two terms in the Karhunen-Loeve expansion  (\ref{KL}).  If in addition that the first eigenfunction is nearly linear in time and explains over 80\% of the total variation of the data, then a parametric linear mixed-effects model with a  linear  time trend for random effects likely will  fit this data well.  This underscores the advantages to use a nonparametric approach such as FDA prior to a model-based longitudinal data analysis for data exploration.  The exploratory analysis  then may suggest viable parametric models that are more parsimonious than FPCA. 


\section{Correlation and  Regression: Inverse Problems and Dimension Reduction for Functional Data}

 As mentioned in Section 1, a major challenge in FDA  is the inverse problem, which  stems from the compactness of the covariance operator.  Consider $\Sigma= \mbox{cov} (X(s), X(t))$ with eigenvalues  $\lambda_k$, $k=1, \ldots, \infty$. If there are only finitely many, say $K$, positive eigenvalues, the functional data become finite dimensional and can be fully described by the   $K$-dimensional principal component scores (in addition to the mean function and the $K$ eigenfunctions).   In this case, the functional data may be viewed as  $K$-dimensional multivariate data and the covariance matrix is equivalent (or isomorphic) to a $K$ by $K$ invertible matrix. 
 Then the same considerations as for  multivariate data apply  and the  inverse problem is equivalent to that of multivariate data with no additional complications for functional data.
 
  If on the other hand that there are infinitely many nonzero, hence positive, eigenvalues, then the covariance operator is a one-to one function. 
 In this case, the inverse operator of $\Sigma$ exists but is an unbounded operator and the range space of the covariance operator is a compact set in $L^2$. This creates a problem to define a bijection, as the inverse of $\Sigma$ is not defined on the entire $L^2$ space.  Therefore regularization is routinely adopted for any procedure that involves an inverse operator. 
 Examples where inverse operators are central include regression and correlation measures for functional data, as $\Sigma^{-1}$ appears in these methods. This inverse problem was for example addressed for functional canonical correlation in \citet{HeMW00} and \citet{HeMW03}, where a solution was proposed under certain constraints on the decay rate of the eigenvalues and the cross covariance operator. 

  \subsection{Functional Correlation}
Different functional correlation measures have been discussed in the literature. Functional Canonical Correlation Analysis serves here to demonstrate some of the problems that one encounters in FDA as a consequence of the non-invertibility of compact operators. 

\medskip
\noi  \textbf{Functional Canonical Correlation Analysis (FCCA).}
Let $(X, Y)$ be a pair of random functions in $L^2(I_X)$ and $L^2(I_Y)$ respectively. The first functional canonical correlation coefficient $\rho_1$ and its associated weight functions $(u_1, v_1)$ are defined as follows, using the 
notation $\langle f_1, f_2 \rangle = \int_I f_1(t)f_2(t)dt$ for any $f_1, f_2 \in L^2(I)$, 
\be
\rho_1= \sup_ {u\in L^2(I_X) , v\in L^2(I_Y)} \cov (\langle u, X \rangle, \langle v, Y \rangle) = \mbox{cov}( \langle u_1, X \rangle, \langle v_1, Y \rangle), \label{CCA1}
\ee
subject to var$(\langle u, X \rangle)=1$ and var$( \langle v, Y \rangle)=1$. Analogously for the  $k$th, $k>1$,  canonical correlation $\rho_k$ and its associated weight functions $(u_k, v_k)$, 
\be
\rho_k= \sup_ {u\in L^2(I_X) , v\in L^2(I_Y)} \mbox{cov} (\langle u, X \rangle, \langle v, Y \rangle) = \mbox{cov}( \langle  u_k, X \rangle,  \langle v_k, Y \rangle), \label{CCAk}
\ee
subject to var$(\langle  u, X \rangle )=1$, var$( \langle  v, Y \rangle)=1$, and that the pair $ (U_k, V_k)= (\langle u_k, X \rangle, \langle v_k, Y \rangle)$ is 
 uncorrelated to all previous pairs $(U_j, V_j)= (\langle  u_j, X \rangle, \langle v_j, Y \rangle)$, for $j=1, \ldots, k-1$.  
 
 Thus, FCCA aims at finding  projections in directions $u_k$ of $X$ and  $v_k$ of $Y$  such that their linear combinations (inner products) $U_k$ and  $V_k$ are maximally correlated, resulting in the series of functional canonical components $(\rho_k, u_k, v_k, U_k, V_k),\, k \ge 1$,  directly extending canonical correlations for multivariate data. Because of the flexibility in the direction $u_1$, which is infinite dimensional, over fitting may occur if the number of sample curves is not large enough. Formally, this is due to the fact that FCCA is an ill-posed problem. 
Introducing the  cross-covariance operator $\Sigma_{XY} : L^2(I_Y)  \rightarrow L^2(I_X)$,
\be  \label{covop}
\Sigma_{XY} v(t) = \int  \mbox{cov} \ (X(t), Y(s)) v(s) ds, 
\ee
for $v \in L^2(I_Y)$ and analogously  the covariance operators for $X$, $\Sigma_{XX}$,  for $Y$,  $\Sigma_{YY}$, and using  $\cov(\langle u, X \rangle , \langle v, Y \rangle)= \langle u, \Sigma_{XY} Y \rangle $, the $k$th canonical component in (\ref{CCAk}) can be expressed as 
\be
\rho_k= \sup_{u\in L^2(I_X),   \langle u, R_{XX}u \rangle=1,  v\in L^2(I_Y), \langle v, R_{YY} v \rangle =1 } \langle u, R_{XY} v \rangle  = \langle u_k, R_{XY} v_k \rangle. \label{CCAk1}
\ee

Then  (\ref{CCAk1})  is equivalent to an eigenanalysis of the operator $R=\Sigma_{XX}^{-1/2} \Sigma_{XY} \Sigma_{YY}^{-1/2}$. Existence of the canonical components is guaranteed if  the operator $R$ is compact. However, the inverse of a covariance operator and the inverses of $\Sigma_{XX}^{1/2}$  or $\Sigma_{YY}^{1/2}$ are not bounded since a covariance operator is compact  under the assumption that the covariance function is square integrable. A possible approach \citep{HeMW03} is to restrict the domain of the inverse to the range $A_X$ of $\Sigma_{XX}^{1/2}$  so that the inverse of $\Sigma_{XX}^{1/2}$ can be defined on $A_X$ and  is a bijective mapping  $A_X$ to $B_X$,  
under some conditions (e.g., Conditions 4.1 and 4.5 in \citet{HeMW03}) on the decay rates of the eigenvalues of $\Sigma_{XX}$ and $\Sigma_{YY}$ and the cross-covariance.   Under those assumptions  the canonical correlations and weight functions  are well defined and exist. 

 An alternative way to get around the above ill-posed  problem is to restrict the maximization in (\ref{CCA1}) and (\ref{CCAk}) to  discrete $l^2$  spaces that are restricted to a reproducing kernel Hilbert space instead of  working within the entire $L^2$ space \citep{EubaH08}. 
Since FCCA is inherently regularized, the first canonical correlation often tends to be too large and its value is difficult to interpret, as it is highly dependent on the value of the regularization parameter.  This overfitting problem, which also can be viewed as a consequence of the high-dimensionality of the weight function, was already illustrated in \citet{LeurMS93}, who were the first to explore penalized FCCA.  functional canonical correlation regularized by a penalty.  
Despite the challenge with overfitting  FCCA can be also employed to implement functional regression problem  by using the canonical weight functions $u_k$, and $v_k$ as bases to expand the regression \cp{HeMW00, HeMW10} .   

Another difficulty with the versions of FCCA  proposed so far is that it requires densely recorded functional data so the inner products in   (\ref{CCAk})  can be evaluated with high accuracy.   Although it is possible to impute sparsely observed functional data using the Karhunen-Lo\'{e}ve expansion (\ref{KL}) before applying any of the  canonical correlations,  a prediction error  will result from such an imputation leading to a biased correlation.  This  bias may be small in practice but  finding an effective FCCA for sparsely observed functional data is still of interest and remains an open problem.    \vs\vs

\noi  \textbf{Other Functional Correlation Measures.} \
The regularization problems for FCCA have motivated the study of alternative notions of functional correlation.  These include singular correlation and singular expansions of paired processes $(X,Y)$. While the first correlation coefficient in FCCA can be viewed as $\rho_{\text{FCCA}}=\sup_{\|u\|=\|v\|=1} \corr(\langle u,X \rangle, \langle v,Y \rangle),$, observing that it is the correlation that induces the inverse problem, one could simply replace the correlation by covariance, i.e., obtain project  functions $u_1, v_1$ that attain $\sup_{\|u\|=\|v\|=1} \cov(\langle u,X \rangle, \langle v,Y \rangle)$. Functions $u_1, v_1$ turn out to be the first pair of the singular basis of 
the covariance operator of $(X,Y)$ \cp{mull:11:2}.
This motivates to define a functional correlation as the first singular correlation 
\be \label{scor}
\rho_{\text{SCA}}=\frac{ \cov(\langle u_1,X \rangle, \langle v_1,Y
\rangle)}{\sqrt{\mbox{var}(\langle u_1,X\rangle)\,\mbox{var}(\langle
v_1,Y\rangle)}}. \ee

Another natural approach that also avoids the inverse problem is to define functional correlation as the cosine of the angle between functions in $L^2$. For this notion to be a meaningful measure of alignment of shapes, one first needs to subtract the integrals of the functions, i.e., their projections on the constant function 1, which corresponds to a ``static part''. Again considering pairs of processes $(X,Y)=(X_1, X_2)$ and denoting the projections on the constant function 1 by  $M_k=\langle X_k, 1 \rangle, \, k=1,2,$   the 
remainder $X_k-M_k,\, k=1,2,$ is the ``dynamic part'' for each random function.  The $L^2$-angle of these dynamic part can thus serves as a distance measure of them, which leads to a correlation measure of functional shapes.  
These ideas can be formalized as follows \cp{mull:05:2}. 
Defining standardized curves either by  $X_k^*(t) = (X_k(t) - M_k)/(\int(X_k(t) - M_k)^2 dt)^{1/2}$ or alternatively by also removing $\mu_k=EX_k$,  
$X_k^*(t) = (X_k(t) - M_k - \mu_k(t))/(\int(X_k(t) - M_k - \mu_k(t))^2 dt)^{1/2}$, the cosine of the angle between the standardized functions is 
$ \rho_{k,l} = E \langle X_k^*,X_l^* \rangle.$ The resulting dynamic correlation and other notions of functional correlation can also be extended to obtain 
a precision matrix for functional data. This approach has been developed  by \ci{opge:06}
for the construction  of a graphical networks for gene time course data.

\subsection{Functional Regression}

Functional regression is an active area of research and the approach depends on whether the responses or covariates 
are functional or vector data and include  combinations of  (i) functional responses with functional covariates, (ii) vector responses with functional covariates,  and  (iii) functional responses with vector covariates.  An approach for  (i) was introduced by \citet{RamsD91} who developed the functional linear model (FLM)  (\ref{FLM3}) for this case, where the basic idea already appears in \ci{gren:50}, who derives this as the regression of one Gaussian process on another.  This model can be viewed as an extension of the traditional  multivariate linear model that associates vector responses with vector covariates.    The topic that has been investigated most extensively in the literature is scenario (ii) for  the case where the responses are scalars and the covariates are functions.  
Reviews of  FLMs are \citet{Mull05, mull:11:5} and a  recent review in  \citet{Morr15}. Nonlinear functional regression models will be discussed in Section 5. In the following we give a brief review of the FLM and its variants. 

\medskip
\noi  \textbf{Functional Regression Models with Scalar Response.}  \ The traditional linear model with scalar response $Y\in \mathcal{R}$     
and vector covariate $\mathbf{X}\in \mathcal{R}^p$ can be expressed as   
 \be   \label{LM}
Y =   \beta_0 + \langle \mathbf{X}, \beta \rangle   + e,   
 \ee
using the inner product in Euclidean vector space, where $\beta_0$ and $\beta$ contain the regression coefficients and $e$ is a zero mean finite variance random error (noise).
Replacing the vector $\mathbf{X}$ in (\ref{LM}) and the coefficient vector $\beta$ by a centered functional covariate  $X^c=X(t)-\mu(t)$ and coefficient function $\beta=\beta(t)$,  for $t \in I$,  one arrives  at the  functional linear model
\be  \label{FLM}
Y =  \beta_0  +  \langle  X^c, \beta \rangle   + e = \beta_0  + \int_I X^c(t) \beta(t) dt +e,   
\ee 
which has been studied  extensively \cp{CardFS99,  CardFS03, hall:07:1}. 

An ad hoc  approach is to expand the covariate $X$ and the coefficient function $\beta$ in the same functional basis, such as the  B-spline basis or 
eigenbasis 
in (\ref{KL}). Specifically, consider an orthonormal basis $\varphi_k, \, k \ge 1,$  of the function space. 
Then expanding both $X$ and $\beta$ in this basis 
leads to   
$X(t)= \sum_{k=1}^{\infty}  A_k \varphi_k (t)$,  $\beta(t)= \sum_{i=1}^{\infty}  \beta_k \varphi_k(t)$ and model (\ref{FLM}) is equivalent to the traditional linear model  (\ref{LM}) of the form
\be  \label{LM2}
Y= \beta_0 + \sum_{k=1}^{\infty} \beta_k A_k    + e,
\ee
where in implementations the  sum on the r.h.s. is replaced by a finite sum that is truncated at the first $K$ terms, in analogy to (\ref{KLK}). 


To obtain consistency for the estimation of  the parameter function $\beta (t)$, $K=K_n$ in (\ref{LM2})  needs to increase with the sample size $n$. For the theoretical analysis, the method of sieves \citep{Gren81} can be applied, where  the $K$th sieve space is the  linear subspace spanned by the first $K=K_n$ components. 
 In addition to the basis-expansion approach, a penalized approach using either P-splines  or smoothing splines has also been studied \citep{CardFS03}.  
  For the special case where  the basis functions $\varphi_k$ are selected as the eigenfunctions $\phi_k$ of $X$, the basis representation approach in (\ref{LM}) is equivalent to conducting a principal component regression albeit with an increasing number of principal components. In this case, however,  the basis functions are estimated rather than pre-specified, and this adds an additional twist to the theoretical analysis.

The simple functional linear model (\ref{FLM}) can be extended to multiple functional covariates $X_1, \ldots, X_p$, also including  additional vector covariates $\mathbf{Z}=(Z_1, \ldots, Z_q)$, where $Z_1=1$,  by 
\be   \label{FLM2}
Y= \langle \mathbf{Z}, \theta  \rangle +  \sum_{j=1}^p \int_{I_j} X^c_j(t) \beta_j(t) dt +e,
\ee
where $I_j$ is the interval where $X_j$ is defined.  In theory, these intervals need not  be the same. 
 Although model (\ref{FLM2}) is a straightforward extension of (\ref{FLM}), its inference is different due to the presence of the parametric component $\theta$. A combined least squares method to estimate  $\theta$ and $\beta_j$ simultaneously in a one step or  profile approach  \citep{HuWC04}, where one estimates $\theta$ by profiling out the nonparametric components $\beta_j$,  is generally preferred  over an alternative  back-fitting method. 
 Once the parameter $\theta$ has been estimated, any approach that is suitable and consistent for fitting the functional linear model (\ref{FLM})  can easily be extended to estimate the nonparametric components $\beta_k$ by applying it to the residuals $Y- \langle \hat{\theta},  \mathbf{Z}  \rangle $. 
 
Extending the  linear setting with a single  index $\int_{I} X^c(t) \beta(t) dt$  to summarize each function covariate, a nonlinear link function $g$ can be added  in (\ref{FLM})  
to create a functional generalized linear model (either within the exponential family or a quasi-likelihood framework and a suitable variance function) 
\be  \label{FGLM}
Y =   g (\beta_0 + \int_I X^c(t) \beta(t) dt )+e. 
\ee 
This model has been considered when  $g$ is known  \citep{jame:02, CardFS03, CardS05, WangQC10} and when it is unknown \cp{MullS05, mull:11:6}. 
When $g$ is unknown and the variance function plays no role, the special case of a  single-index model 
has further been extended to multiple indices, the number of which is possibly unknown.  Such ``multiple functional index models'' typically forgo the additive error structure imposed in (\ref{FLM}) - (\ref{FGLM}), 
\be  \label{FIM}
Y =   g(\int_I   X^c(t) \beta_1(t) dt,  \ldots,  \int_I   X^c(t) \beta_p(t) dt , e),
\ee  
where $g$ is an unknown multivariate function on $\mathcal{R}^{p+1}$.  
This line of research follows the paradigm of sufficient  dimension reduction approaches, which was first proposed  for vector covariates as an off-shoot of sliced inverse regression (SIR) \citep{DuanL91, Li91}, which has been extended to functional data in \citet{FerrY03, FerrY05, CookFY10} and to longitudinal data in \citet{JianYW14}. 


\medskip

\noi  \textbf{Functional Regression Models with Functional Response.}  For  a function $Y$  on $I_Y$ and a single functional covariate $X(t), \, s\in I_X$, two major models have been considered,
\be  \label{VC}
Y(s)= \beta_0(s) + \beta(s) X(s) +e(s),
\ee
and 
\be \label{FLM3}
Y(s)= \alpha_0(s) +   \int_{I_X}  \alpha(s, t) X^c(t) dt +e(s), \ee
where $\beta_0(s)$ and $\alpha_0(s)$ are non-random functions that play the role of functional intercepts,  and $\beta(s)$ and $\alpha(s,t)$ are non-random coefficient functions that play the role of functional  slopes. 

Model  (\ref{VC}) implicitly assumes that $I_X=I_Y$  and is most often referred to as  ``varying-coefficient''  model.  Given $s$, $Y(s)$ and $X(s)$ follow the traditional linear model, but  the covariate effects may change with time $s$.   This model assumes that the value of $Y$ at time $s$ depends only on the current value  of $X(s)$ and not the history $\{X(t): \, t \leq s \}$ or future values, hence it is a ``concurrent regression model''.  A simple and effective  approach to estimate $\beta$ is to first fit model (\ref{VC}) locally in  a neighborhood of  $s$ using ordinary least square methods to get an initial estimate $\tilde{\beta}(s)$, and then to smooth these initial estimates $\tilde{\beta}(s)$ across $s$ to get the final estimate $\hat{\beta}$ \citep{FanZ99}. In addition to such a two-step procedure, one-step smoothing methods  have been also studied \citep{HoovRWY98, WuC00, ChiaRW01, EggeEL10, HuanWZ02}, as well as 
hypothesis testing and confidence bands \cp{WuCH98, HuanWZ04}.
There are also two review papers \citep{WuY02, FanZ08}  on this topic.  More complex varying coefficient models include the nested model in \citet{BrumR98} ,  the covariate adjusted model in \citet{SentM05}, and the multivariate varying-coefficent model in  \citet{ZhuFK14}, among others. 

Model (\ref{FLM3}) is generally referred to as functional linear model (FLM), and it differs in crucial aspects from the varying coefficient model  (\ref{VC}): At any given time $s$, the value of $Y(s)$ depends on the entire trajectory of $X$.  It  is a direct  extension of traditional linear models with multivariate response and vector covariates by changing the inner product from the Euclidean vector space to $L^2$.  This model also is a  direct extension of model  (\ref{FLM}) when the scalar $Y$ is replaced by  $Y(s)$ and the coefficient function $\beta$  varies  with $s$, leading to a bivariate coefficient surface. It was first studied by \citet{RamsD91}, who proposed a penalized least squares method to estimate the regression coefficient surface $\beta(s, t)$.  When $I_X=I_Y$, it is often reasonable to assume that only the history of $X$ affects $Y$, i.e., that $\beta(s,t)=0$ for $s < t$.  This has been referred to as the 
``historical functional linear model'' \cp{malf:03}, because only the history of the covariate is used to model the response process. This model deserves more attention. 


When $X \in \mathcal{R}^p$ and $Y \in \mathcal{R}^q$ 
are random vectors,  the normal equation of the least squares regression of $Y$ on $X$ is cov$(X, Y)=$cov$(X, X) \beta$, where  $\beta$ is a $p\times q$ matrix.   Here a solution can be easily obtained if cov$(X, X)$ is of full rank so its inverse exists.  An extension of the normal equation to functional $X$ and $Y$ is straightforward  by replacing the covariance matrices by their corresponding covariance operator.    However,  an ill-posed problem emerges for the functional normal equations.  Specifically, if for paired processes $(X,Y)$ the cross-covariance function is
$r_{XY}(s, t)= \mbox{cov}(X(s), Y(t))$  and $r_{XX}(s,t)=\mbox{cov}(X(s), X(t))$ is the auto-covariance function of $X$, we define the linear operator, $R_{XX}: L^2 \times L^2 \rightarrow L^2  \times L^2$ by $(R_{XX} \beta) (s, t) = \int r_{XX} (s, w) \beta (w, t) dw.$ Then a ``functional normal equation'' takes the form \cp{HeMW00} $$r_{XY}= R_{XX} \beta, \  \mbox{for} \ \beta \in L^2(I_X  \times I_X).$$  Since $R_{XX}$ is a compact operator in $L^2$, its inverse is not bounded leading to an ill-posed problem.   Regularization is thus needed in analogy to the situation for FCCA described in Section 3.1 \cp{HeMW03}.  
The functional linear model (\ref{FLM}) is similarly ill-posed but not for the varying coefficient model (\ref{VC}) because the normal equation for the varying-coefficient model can be solved locally at each time point and does not involve inverting an operator.  

Due to the ill-posed nature of the functional linear model,   the asymptotic behavior of the regression estimators varies in the three design settings.  For instance,  
 a $\sqrt{n}$ rate is attainable under the varying-coefficient model (\ref{VC}) for completely observed functional data or dense functional data possibly contaminated with measurement errors,  but not for the other two functional linear models (\ref{FLM}) and (\ref{FLM3}) unless the functional data can be expanded by a finite number of basis functions.  The convergence rate for (\ref{FLM}) depends on how fast the eigenvalues decay to zero and on regularity assumptions on $\beta$ \cp{cai:06,hall:07:1}  even when functional data are observed continuously without error.  An interesting phenomenon is that prediction for model (\ref{FLM}) follows a different paradigm in which  $\sqrt{n}$ convergence is attainable if the predictor $X$ is sufficiently smooth and the eigenvalues of predictor processes are well behaved \citep{cai:06}.  Estimation for $\beta$ and asymptotic theory for model (\ref{FLM3}) were explored in \citet{YaoMW05b, HeMW10} for sparse functional data.  

As with scalar responses, both the varying coefficient model (\ref{VC}) and functional linear model (\ref{FLM3}) can accommodate vector covariates   and multiple functional covariates.  Since each component of the vector covariate can be treated as a functional covariate with a constant value  we only  discuss  the extension to multiple functional covariates, $X_1, \ldots, X_p$, noting that interaction terms can be added as needed.  The only change we need to make on the models is to replace the  term $\beta(s)X(s)$ in (\ref{VC}) by $\sum_{j=1}^p \beta_j(s) X_j(s)$ and  the term   $\int_{I_X}  \beta(s, t) X(t) dt $  in (\ref{FLM3})  by $\sum_{j=1}^p \int_{I_{X_j}}  \beta_j(s, t) X_j(t) dt$, where $I_{X_j}$ is the domain of $X_j$.  If there are many predictors, a variable selection problem may be encountered, and when using basis expansions it is natural to employ a group lasso or similar constrained multiple variable selection method under sparsity or other suitable assumptions. 

Generalized versions can be developed by adding a pre-specified  link function $g$ in models  (\ref{VC}) and (\ref{FLM3}). For the case of the varying coefficient model 
and  sparse functional data  this has been investigated in \citet{SentM08} for the generalized varying coefficient model and for  model (\ref{FLM3}) and dense functional data in \citet{jame:05} for a finite number of expansion coefficients for each function. 
 \citet{JianW11} considered  
a setting where the link function may vary with time but the $\beta$ in the index does not change vales overtime .  
The proposed dimension reduction approach  expands the MAVE method by \citet{XiaTL02} to functional data. \vspace{.3cm}

\noi \textbf{Random Effects Models}. 
In addition to targeting fixed effects regression, the nonparametric modeling of random effects is also of interest. One approach is to extend the FPCA approach of Section 2 to   incorporate covariates \cp{CardS06, JianAW09,JianW10}. 
These approaches are aiming to incorporate  low dimensional projections of covariates to alleviate the curse of dimensionality for nonparametric procedures. One scenario where it is easy to implement covariate adjusted FPCA is the case where one has functional responses and vector covariates. One could  conduct a  pooled FPCA combining all data as a first step and then to use the FPCA scores obtained from the first stage to model covariate effects through a single-index model at each FPCA component \cp{ChioMW03a}. At this time, such approaches require dense functional data,  as for sparse data individual FPC scores cannot be estimated consistently.


\bco

  -{\color{red} Testing in functional linear model includes \citet{ShenF04, SweiGC13}  and \citet{HilgMV13}.  Check if they only consider linear link function. Mention that nonlinear is an open problem...}

 \citet{ShenF04} extended the ANOVA test for coefficients in  conventional linear model  to testing the coefficient function in functional linear model. 
 
\fi

\section{Clustering and classification of functional data} \label{sec:cluster.class}

Clustering and classification are useful tools for traditional multivariate data analysis and are equally important yet more challenging in functional data analysis. Clustering aims to group a set of data into a configuration in such a way that data objects within clusters are more similar than across clusters with respect to a  certain metric.  In contrast, classification aims to assign an individual to a pre-determined group or class based on class-labeled observations. 
In the terminology of machine learning, functional data clustering is an unsupervised learning process while functional data classification is a supervised learning procedure. 
While clustering aims to identify groups using a clustering criterion, classification assigns a new data object to a pre-determined group by a discriminant function or a classifier. Functional classification typically involves training data containing a functional predictor with an associated multi-class label for each data object.  
The discrimination procedure of functional classification is closely related to functional clustering, although the goals are different.  When cluster centers can be established in functional data clustering, the criteria for finding clusters can also be used for classification.   
Methodology for  clustering and classification of functional data has advanced rapidly during the past decades, due to rising demand for such methods in data applications. In view of the vast literature on functional clustering and classification, we focus in the following on only a few typical methods. 

\subsection{Clustering of functional data}

For vector-valued multivariate data, hierarchical clustering and the $k$-means methods are two classical and popular approaches. Hierarchical clustering is an algorithmic approach, using either agglomerative or divisive strategies, that requires a dissimilarity measure between sets of observations to decide which clusters should be combined or where a cluster should be split. In the $k$-means clustering method, the underlying assumption hinges on cluster centers, the means of the clusters. The cluster centers are defined through algorithms aiming to partition the observations into $k$ clusters such that the within-cluster sum of squares, centering around the means, is minimized.  
Classical clustering concepts for vector-valued multivariate data can typically be extended to functional data, where various additional considerations arise, such as discrete approximations of distance measures, and dimension reduction of the infinite-dimensional functional data objects. 
In particular,  $k$-means type clustering algorithms have been widely applied to functional data, and  
are more popular than hierarchical clustering algorithms. It is  natural to view cluster mean functions as the cluster centers in functional clustering. 

Specifically, for  a sample of functional data $\{X_i(t); \, i=1,\ldots, n\}$, the  $k$-means functional clustering aims to find a set of cluster centers $\{\mu^{1}, \ldots,\mu^{L}\}$, assuming there are $L$ clusters, by minimizing the sum of the squared distances between $\{X_i\}$ and the cluster centers that are associated with their cluster labels $\{C_i; \, i=1,\ldots, n\}$, for a suitable functional distance $d$. That is, the $n$ observations $\{X_i\}$ are partitioned into $L$ groups such that  
\begin{equation} \label{k.means}
 \frac{1}{n} \sum_{i=1}^{n} d^2(X_i, \mu_{n}^{c}), 
\end{equation}
is minimized over all possible sets of functions $\{\mu_{n}^{c};\  c=1, \ldots,L\}$, where $\mu_{n}^{c}(t)=\sum_{i=1}^n X_i(t) \mb{1}_{\{C_i=c\}}/N_c$, and $N_c=\sum_{i=1}^n \mb{1}_{\{C_i=c\}}$. 
Since functional data are discretely recorded, frequently contaminated with measurement errors, and can be sparsely or irregularly sampled, a common approach to achieve~\eqref{k.means} is to project functional data of infinite-dimension onto a low dimensional space of a set of basis functions, similarly to the implementations of functional correlation and regression.  The 
distance $d$ is often chosen as  the $L^2$ distance. The following two basic approaches are commonly used in functional data clustering.\vs\vs

\noindent\textbf{Functional Basis Expansion Approach.} 
As described before, once a set of basis functions $\{\varphi_1, \varphi_2, \ldots \}$ of $L^2$ has been chosen, the first $K$ projections $B_k=\langle X^c, \varphi_k \rangle, \, \, k=1, \ldots, K,$  of the observed  trajectories onto the space spanned by the  basis functions are used to represent the functional data. 
An example is the truncated Karhunen-Lo{\`e}ve expansion of $X_i$ in (\ref{KLK}), where the basis is chosen as the eigenfunctions 
 $\{\phi_1, \phi_2, \ldots \}$ of the auto-covariance operator of the underlying process $X$.   
\bco
If we consider a set of basis functions  then the Bessel's inequality implies that, for any function $X$ in $L^2$, the infinite sum 
$\sum_{k=1}^{\infty} \langle X, \varphi_k \rangle \varphi_k(t)$ 
converges to $X(t)$. Therefore, for a sample of curves $X_i(t)$, we may consider the truncated functional basis expansion model to approximate $X_i(t)$,
\begin{equation} \label{eq:basis.exp}
\tilde{X}_i(t)=\sum_{k=1}^{L_{B}} B_{ik} \varphi_k(t), 
\end{equation}
where the basis coefficients $B_{ik}$ are unknown parameters to be fitted, and the number of components $L_{B}$ needs to be determined. 
While the set of the basis functions $\{\varphi_k\}$ in~\eqref{eq:basis.exp} is chosen subjectively, the set of basis functions $\{\phi_k\}$ in~\eqref{eq:fpca.exp} is determined data-adaptively by the covariance operator, and the components are ordered in the sense that the variance of $\{A_{ik}\}$ corresponding to $\{\phi_k\}$ are descending in $k$. 
Similarly to the role of the FPC scores $\{A_{ik}; k=1, \ldots, L_{A}\}$, the set of the basis coefficients $\{B_{ik}; k=1, \ldots, L_{B}\}$ serves as the proxies of the functional observation $X_i$ in $L^2$. The expressions of functional data using~\eqref{eq:fpca.exp} and~\eqref{eq:basis.exp} play the central roles in further development of functional clustering methods as evidenced by a large body of the literature. 
\fi
Since functional data are realizations of random functions, it is intuitive  to consider the stochastic structure of random functions to determine the cluster centers
\bco . In particular, the mean function and the set of eigenbases in the Karhunen-Lo{\`e}ve expansion of a random function comprise a subspace in $L^2$, which determines the structure of the random function.  It is sensible to consider cluster centers \fi
as the subspaces spanned by the mean and the set of the eigenfunctions, the structure of the random functions, in contrast to the $k$-means functional clustering that takes the mean functions as cluster centers. This idea is implemented in the subspace projected functional data clustering approach \cp{chio:07, chio:08}. Moreover, statistical models can also be used as cluster centers to depict a group of similar data, for example by applying  mixture models.


There is a vast amount of the literature on functional data clustering during the past decade, including methodological development and a broad range of applications. Some selected approaches  to be discussed below include  the
$k$-means type of clustering in Section~\ref{subsec:mean}.  This is followed in  Section~\ref{subsec:subspace}  by more details about the subspace projected clustering methods, and in  Section~\ref{subsec:model} by  model-based functional clustering approaches. 

\subsubsection{Mean functions as cluster centers} \label{subsec:mean} 
The traditional $k$-mean clustering  for vector-valued multivariate data has been extended to  functional data using the mean function  as cluster centers.  
The approaches can be divided into two categories.\vs

\noindent\textbf{Functional Clustering via Functional Basis Expansion.} \ 
In the functional basis expansion approach, the  functional data are projected onto the same set of basis functions irrespective of cluster membership, and the sets of basis coefficients $\{B_{ki};\, k=1,\ldots,K\}$ serve as the proxies of individual trajectories. Thus, the distribution patterns of the $\{B_{ik}\}$ reflect the clustering patterns of the set of functional data.  A typical functional clustering approach then is to represent  the functional data by the coefficients of a  basis expansion,  this requires carefully choosing a particular set of basis functions, and then using available clustering algorithms for multivariate data, such as the $k$-mean algorithm, to partition the estimated sets of coefficients.   

Such two stage clustering has been adopted in~\cite{Abraham.etal(2003)} using B-spline basis functions and \cite{Serban.Wasserman(2005)} using Fourier basis functions coupled with the $k$-means algorithm,  as well as \cite{Garcia-Escudero.Gordaliza(2005)} using B-splines with a robust trimmed $k$-means method.  
By clustering the fitted sets of coefficients $\{B_{ik}\}$ through the $k$-means algorithms, one obtains the set of cluster centers $\{\bar{B}_{1}^{c}, \ldots, \bar{B}_{K}^{c}\}$ on the projected space, and thus the set of cluster centers $\{\mu^{c} ; \, c=1, \ldots, L\}$, where  
$
\mu^{c}(t) =\sum_{k=1}^{K} \bar{B}_{k}^{c} \varphi_k(t). 
$
\cite{Abraham.etal(2003)} derived the strong consistency property of this clustering method that has been implemented with various basis functions, such as 
%
P-splines~\citep{Coffey.etal(2014)},  a 
Gaussian ortho-normalized basis~\citep{Kayano.etal(2010)}, and the wavelet basis~\citep{Giacofci.etal(2013)}.\vs 

\noindent\textbf{Functional Clustering via FPCA.} \ 
In contrast to the functional basis expansion approach that need to choose a particular set of basis functions, the finite approximation FPCA approach (\ref{KLK}) uses data-adaptive basis functions that are determined by the covariance function of the functional data.   
Then the distributions  of the sets of FPCs  $\{A_{ik}\}$ indicate different cluster patterns, while the overall mean function $\mu(t)$ does not affect clustering, and the scores $\{A_{ik}\}$ play a similar role as the basis coefficients $\{B_{ik}\}$ for clustering. 
\cite{mull:08:8} used a $k$-means algorithm on the FPCs,  employing a special distance adapted to clustering sparse functional data,   and \cite{chio:07} used a $k$-means algorithm on the FPCs  as an initial clustering step for the subspace projected $k$-centers functional clustering algorithm.  When the mean functions are the cluster centers, the initial step of the approach works reasonably well. However, when the cluster centers reflect  specific features of the structure of the covariance functions, the approach of \cite{chio:07} to be described in the next subsection can further improve the quality of clustering.

\subsubsection{Subspaces as cluster centers.} \label{subsec:subspace}

Clusters can be defined via subspace projection such that cluster centers lie on sets of basis functions of cluster subspaces, rather than mean functions.  This idea is particularly sensible in functional data clustering by observing that the truncated Karhunen-Lo{\`e}ve representation (\ref{KLK}) of a random function in $L^2$ comprises a fixed component, a mean function, and a random component, a linear combination of the eigenfunctions of the covariance operator with weights determined by the FPCs. Since each cluster contains a subset of data sampled from random functions in $L^2$, and  each subset of data lies in a subspace of $L^2$, the structure of the stochastic representation can be used to identify clusters of functional data.  \cite{chio:07} consider a FPC subspace spanned by  a mean function and a set of eigenfunctions, and define clusters via FPC subspace projection. The ideas are briefly explained as follows.

Let $C$ be the cluster membership variable, 
 and the FPC subspace $\mathcal{S}^{c}=\{ \mu^{c}, \phi_{1}^{c}, \ldots, \phi_{K_c}^{c}\}$, $c=1, \ldots, L$, assuming that there are $L$ clusters.  The projected function of $X_i$ onto the FPC subspace $\mathcal{S}^{c}$ can be written as
\begin{equation}
\tilde{X}_{i}^{c}(t)=\mu^{c}(t)+\sum_{k=1}^{K_c} A_{ik}^{c} \phi_{k}^{c}(t) . 
\end{equation}
The \textit{subspace-projected $k$-centers functional clustering procedure}~\citep{chio:07} aims to find the set of cluster centers $\{\mathcal{S}^{c}; \, c=1, \ldots, K\}$, such that the best cluster membership of $X_i$, $c^*(X_i)$, is determined by minimizing the discrepancy between the projected function $\tilde{X}_i^{c}$ and the observation $X_i$ such that 
\begin{equation} \label{eq:k.centers}
c^{*}(X_i)=\argmin_{c\in\{1, \ldots, L\}} \sum_{i=1}^{n} d^2( X_i ,\tilde{X}_{i}^{c}).
\end{equation}

In contrast, the $k$-means clustering aims to find the set of cluster sample means as the cluster centers, rather than the subspaces spanned by $\{\mathcal{S}^{c}; c=1, \ldots, L\}$ as the cluster centers. The initial step of the subspace-projected clustering procedure considers that $\mathcal{S}^{c}$ contains only $\mu^{c}$, which reduces to the $k$-means functional clustering. In the iterative steps, the set of eigenfunctions for each cluster is obtained and  identifies the set of cluster subspaces $\{\mathcal{S}^{c}\}$. The iteration runs until convergence. This functional clustering approach simultaneously identifies the structural components of the stochastic representation for each cluster. The idea of the $k$-centers function clustering via subspace projection was further developed to clustering functional data with similar shapes based on a shape function model with random scaling effects~\citep{chio:08}. 

More generally, in probabilistic clustering the cluster membership of $X_i$ may be determined by maximizing the conditional cluster membership probability given $X_i$, $P_{C\mid X}(c\mid X_i)$, such that
\begin{equation} \label{eq:criterion.C.X}
c^*(X_i)=\argmax_{c\in\{1, \ldots, L\}} P_{C\mid X}(c\mid X_i).
\end{equation} 
This criterion requires modeling of the conditional probability $P_{C\mid X}(\cdot\mid \cdot)$. It can be achieved by a generative approach that requires a joint probability model or alternatively through a discriminative approach using, for example, a multi-class logit model~\citep{Chiou(2012)}. 

For the $k$-means type or the $k$-centers functional clustering algorithms, the number of clusters is pre-determined. The number of clusters for subspace projected functional clustering can be determined by finding the maximum number of clusters while retaining significant differences between pairs of cluster subspaces.~\citet{Li.Chiou(2011)} developed the forward functional testing procedure to identify the total number of clusters under the framework of subspace projected functional data clustering. 

\subsubsection{Mixture models as the cluster centers} \label{subsec:model}

Model-based clustering \citep{Banfield.Raftery(1993)} using mixture models is widely used in clustering vector-valued multivariate data and has been extended to  functional data clustering. Here  the models of the mixtures underlie the cluster centers. Similarly to the $k$-means type of functional data clustering, model-based approaches to functional data clustering  start by  projecting infinite dimensional functional data onto low-dimensional subspaces. E.g., \ci{James.Sugar(2003)} introduced functional clustering models  based on Gaussian mixture distributions for the natural cubic spline basis coefficients, with emphasis on clustering sparsely sampled functional data.  Similarly, ~\cite{Jacques.Preda(2014), Jacques.Preda(2013)} applied the idea of Gaussian mixture modeling to FPCA scores. All these methods are based on truncated expansions as in (\ref{KLK}).

Random effect modeling also provides a  model-based clustering approach, that can be based on  mixed effects models with B-splines or P-splines, for example  to cluster time-course gene expression data \cp{Coffey.etal(2014)}. For clustering longitudinal data, a linear mixed model for clustering using a penalized normal mixture as random effects distribution has been studied \cp{Heinzl.Tutz(2014)}.
Bayesian hierarchical clustering also plays an important role in the development of model-based functional clustering, which typically assumes Gaussian mixture distributions on the sets of basis coefficients fitted to individual trajectories. Dirichlet processes are frequently used as the prior of the mixture distributions and to deal with uncertainty of cluster numbers \citep{Angelini.etal(2012),Rodriguez.etal(2009), Petrone.etal(2009), Heinzl.Tutz(2013)}.  
\bco Further development using Dirichlet processes include the dependent Dirichlet process mixtures of Gaussian distributions \citep{Rodriguez.etal(2009)},  the hybrid Dirichlet mixture models for functional data \citep{Petrone.etal(2009)}, and clustering linear mixed models with approximate Dirichlet process mixtures \citep{Heinzl.Tutz(2013)}. 
\fi

\subsection{Classification of functional data}


While functional clustering aims at finding clusters by minimizing an objective function such as~\eqref{k.means} and~\eqref{eq:k.centers}, or more generally, by maximizing the conditional probability as in~\eqref{eq:criterion.C.X}, functional classification assigns a group membership to a new data object with a discriminant function or a classifier.  
Popular approaches for functional data classification are based on functional regression models that feature  the class labels as the response variable and the observed functional data and other covariates as the predictors. This view leads to the development of regression based functional data classification using, for example, functional generalized linear regression models and functional multiclass logit models. Similar to approaches of functional data clustering, most functional data classification methods  apply a dimension reduction technique using a truncated expansion in a pre-specified function basis or in the data-adaptive eigenbasis. 

\subsubsection{Functional regression for classification}
For regression-based functional classification models, functional generalized linear models \citep{jame:02,Mull05} or more specifically, functional binary regression,  such as functional logistic regression, are popular approaches.  
Let $\{(Z_i, X_i); \, i=1,\ldots,n\}$ be a set of random sample, where $Z_i$ represents a class label, $Z_i\in\{1, \ldots, L\}$ for $L$ classes, associated with the observation $X_i$.  A classification model for an observation $X_0$ based on functional logistic regression is 
\begin{equation} \label{eq:class.reg}
\log\frac{Pr(Z=k\mid X_0)}{Pr(Z_i=L\mid X_0)}= \gamma_{0k} + \int_{\cal T} X_0(t) \gamma_{1k}(t) dt , \quad k=1,\ldots, L-1,
\end{equation}
where $\gamma_{0k}$ is an intercept term and $\gamma_{1k}(t)$ is the coefficient function of the predictor $X_0(t)$ to be fitted by the sample data. Here, $Pr(Z_i=L\mid X_i)= 1-\sum_{k=1}^{L}Pr(Z_i=k\mid X_i)$. This is a functional extension of the baseline odds model in multinomial regression \cp{mccu:83:1}. 

Given a new observation $X_0$, the model-based Bayes classification rule is to choose the class label $Z_0$ with the maximal posterior probability among $\{Pr(Z_0=k\mid X_0); \, k=1, \ldots, L\}$.
More generally, \cite{mull:06:8} used the generalized functional linear regression model based on the FPCA approach. When the logit link is used in the model, it becomes the functional logistic regression model, 
several variants of which have been studied \cp{Araki.etal(2009), Matsui.etal(2011),Wang.etal(2007),  Zhu.etal(2010), Rincon.Ruiz-Medina(2012)}. 
 
 \bco methods were developed based on the logistic regression classification model coupled with the functional basis expansion approaches to deal with the functional predictor. \cite{Araki.etal(2009), Matsui.etal(2011)} adopted regularized functional basis expansion approaches using Gaussian basis functions with hyperparameters estimated via the $k$-means clustering algorithm.  Wavelet basis functions were used in several studies, including the Bayesian curve classification approach of~\citep{Wang.etal(2007),  Zhu.etal(2010)} and the RKHS-based approach to penalized logistic regression of~\cite{Rincon.Ruiz-Medina(2012)}. \fi

\subsubsection{Functional discriminant analysis for classification}
In contrast to the regression-based functional classification approach, another popular approach is based on the classical linear discriminant analysis method.  The basic idea of this approach is to classify according to  the largest conditional probability of the class label variable given a new data object by the Bayes rule or classifier.  Suppose that the $k$th  class has prior probability $\pi_k$, $\sum_{k=1}^K \pi_k=1$. Given the density of the $k$th class, $f_k$, the posterior probability of a new data object $X_0$ is given by the Bayes formula, 
\begin{equation} \label{eq:class.reg}
Pr(Z=k\mid X_0)=\frac{\pi_k f_k(X_0)}{\sum_{j=1}^{K} \pi_j f_j(X_0)} .
\end{equation}

Developments along these lines include 
a functional linear discriminant analysis approach to classify curves    \cp{James.Hastie(2001)},   
a functional data-analytic approach to signal discrimination, using the FPCA method for dimension reduction  \cp{Hall.etal(2001)}  and kernel functional classification rules for  nonparametric curve discrimination
\cp{Ferraty.Vieu(2003), Chang.etal(2014),Zhu.etal(2012)}.


\section{Nonlinear Methods for Functional Data}

Due to the complexity of functional data analysis, which blends stochastic process theory, functional analysis, smoothing and multivariate techniques, most research at this point has focused on
 linear functional models, such as functional principal components and functional linear regression \cp{rams:05,cai:06,hall:07:1,mull:08:3,ritz:09}. Perhaps owing to the success of  these linear approaches, the development of nonlinear methods has been much slower. However, in many situations linear methods are not adequate. A case in point is the presence of time variation or time warping in many data. This means that observation time itself is randomly distorted and sometimes time variation  constitutes the main source of variation \cp{wang:97}. Statistically efficient models will then need to reflect the nonlinear features in the data.

\subsection{Nonlinear Regression Models}

The classical functional regression models are  linear models with a combination of functional and scalar components in predictors and responses. Models with a linear predictor such as the generalized functional linear model and single index models also have usually nonlinear link functions and their analysis is much more complex than that of the functional linear model. Yet they still maintain  many similarities with linear functional models. The boundary between linear and nonlinear models is thus in flux. 

Due to the increased flexibility of nonlinear and nonparametric models for functional data, one needs to walk a fine line in extending functional linear models. 
For example, there have been various developments towards fully nonparametric regression models for functional data \cp{ferr:06}. 
These models extend the concept of nonparametric smoothing to the case of predictor functions, where for scalar responses $Y$ one considers functional predictors $X$, aiming at $E(Y\mid X)=g(X)$ for a smooth regression function $g$.  Such an approach  is motivated by  extending usual smoothing methods, such as kernel smoothers, by replacing all differences in the predictor space by a functional distance, so that 
the scaled kernel $K(\frac{x-y}{h})$ with a bandwidth $h$ becomes  $K(\frac{d(x,y)}{h})$, where $d$ is a metric  in the predictor space. For a comprehensive review of this approach we refer to \ci{ferr:06}. 
Due to the infinite nature of the predictors, in the unrestricted general functional case such models are subject to a serious form of ``curse of dimensionality'', as the
  predictors are inherently infinite-dimensional. Formally, this is due to the infinite-dimensional nature of functional predictors and the associated
unfavorable small ball probabilities in function space \cp{hall:10}. In some cases, when data are clustered in lower-dimensional manifolds, the rates of convergence of the lower dimension will likely apply, as is the case for nonparametric regression \cp{bick:07}, counteracting this curse. 

To avoid the curse from the start, it is of interest to consider more structured nonparametric models, which sensibly balance
sufficient structure with increased flexibility. Structural stability is usually  satisfied if one obtains polynomial rates of convergence of the estimated structural components and of the predictors. A variety of such models have been studied in recent years. 
Popular extensions of classical linear regression include single or multiple index models, additive models and polynomial regression. Analogous extensions of functional 
linear regression models have been studied. Extensions  to single index models \cp{mull:11:6}  provide enhanced flexibility and structural stability with usually polynomial rates of convergence. Beyond single index models, another powerful  dimension reduction tool is  the additive model \citep{Ston85, HastT86}, which has been extended to functional data \cp{LinZ99, YouZ07, CarrMM09, LaiHL12, WangXQ14}.    In these models it is assumed that the time effect is also additive, which may be somewhat  restrictive. \citet{ZhanPW13}  studied a time-varying additive model whose additive components are bivariate functions of time and a covariate.  A downside is that two-dimensional smoothing is needed for each component and that the covariate effect is entangled with the time effect.  A special case of this model where one assumes that each of the additive components is the product of an unknown  time effect and an unknown  covariate effect \cp{ZhanW15} involves only one-dimensional smoothing  and is easy to interpret and implement.       Below we describe a simple additive approach that  exploits the independence of FPCA scores.  \vs

\no {\bf Additive Functional Regression.} Various extensions of additive models to {\it additive functional models} are also of interest. A first option is to utilize functional principal components (s) or scores  $A_{k}$ as defined in (\ref{KL})
for dimension reduction of the predictor process or processes $X$, and then to assume that the regression relation is additive in these, rather than linear. While the linear functional regression model with scalar response can be written as $E(Y \mid X)=EY + \sum_{k=1}^{\infty} A_k \beta_k$ (cf. (\ref{LM2})) with an infinite sequence of regression coefficients $\beta_k$, the extension to the additive model is  the {\it functional additive model} 
\be \la{fam} E(Y \mid X)=EY + \sum_{k=1}^{\infty} f_k(A_k),\ee
where the component functions are required to be smooth and to satisfy $E(f_k(A_k))=0$ \cp{mull:08:2, sood:08}.

This model can be characterized as frequency-additive. A key feature that makes this model not only easy to implement but also accessible to 
asymptotic analysis even when considering infinitely many predictor components, i.e., the entire infinite-dimensional predictor process, is 
a consequence of the observation that  (with $\mu_Y=EY$) 
 \be \la{fam1c}
E(Y-\mu_Y\mid A_{k})=E\{E(Y-\mu_Y|X) \mid A_k\} =E\{\sum_{j=1}^\infty
f_j(A_j) \mid A_k\}=f_k(A_k), \ee
if the functional principal components are assumed to be independent. 
In this case,  simple one-dimensional smoothing of the responses against the 
FPCA scores leads to consistent estimates of the component functions $f_k$ \cp{mull:08:2}.  
A similar phenomenon applies to functional linear model in that  $E(Y-\mu_Y\mid A_{k})=\beta_k A_k$, because of the uncorrelatedness of the FPCA scores of the predictor processes. A consequence of this is that a functional linear regression can be decomposed  into  a sequence of infinitely many simple linear regressions \cp{mull:07:3, mull:09:7}. 


Projections on a finite number of directions 
for each of potentially many predictor functions that are guided by the relationship between predictors and responses provide an alternative additive approach that, while  ignoring the infinite dimensional nature of the predictors, is practically promising since the projections are formed by taking into consideration the relation between $X$ and $Y$, in contrast to other functional regression models where the predictors are formed merely based on the auto-covariance structure of predictor processes $X$ \cp{jame:05,mull:11:6,fan:14}.

Still other forms of additive models have been considered for functional data. While model (\ref{fam}) can be characterized as frequency-additive, as it is additive in the FPCs, one may ask the question whether there are time-additive models. It is immediately clear that since the number of time points on an interval domain  is uncountable, an unrestricted time-additive model $E(Y \mid X)= \sum_{t \in [0,T]} f_t(X(t))$ is not feasible. One can resolve this conundrum by assuming that the functions $f_t$ are smoothly varying in $t$. Then considering a sequence of time-additive models on increasingly dense finite grids of size $p$, 
$$E(Y|X(t_1),\ldots,X(t_p))=\sum_{j=1}^p \,f_j(X(t_j)),$$
assuming
that $f_j(x)=g(t_j,x)$ for a smooth bivariate function $g$, leads in the limit $p \rightarrow \infty$ to the {\it continuously additive model} \cp{mull:13:2}
\be \la{cam} E(Y|X)=\lim_{p\rightarrow \infty} \, \frac{1}{p}\sum_{j=1}^p \,g(t_j,X(t_j))=\int_{[0,T]}\,g(t,X(t))\,dt.\ee 

This model can be implemented with a bivariate spline representation of the function $g$; it was 
discovered independently by \ci{mcle:14}.
Nonlinear or linear models where individual predictor times are better predictors  than functional principal components, i.e., regression models with time-based rather than frequency-based predictors, have also been considered. These models can be viewed as special cases of the continuously additive model (\ref{cam}) in that only a few time points and their associated additive functions $f_j(X(t_j))$ are assumed to be predictive \cp{ferr:10}.\vs

\noi {\bf Optimization and Gradients With Functional Predictors.}  In some applications one may  wish to maximize the response $E(Y \mid X)$ in terms of features of the predictor function $X$. Examples where this is relevant include  the evolution of life history trajectories such as reproductive trajectories $X$ in medflies. Maximization of lifetime reproduction $Y$ provides an evolutionary advantage but must be  gauged against mortality, which cuts off further reproduction and is known to rise under strong early reproduction through the ``cost of reproduction''. Generally, the outcome $Y$ is a characteristic to be maximized. Therefore,  gradients in terms of functional predictors $X$ are of interest. Extending the functional additive model, one can introduce additive gradient operators with arguments in $L^2$ at each predictor level $X \equiv \{A_1, A_2, \ldots\}$,
\be
\la{ga1} \Gamma_{X}^{(1)}(u)=\sum_{k=1}^\infty f_k^{(1)}(A_{k})\int \phi_k(t)u(t)dt,\quad  u \in L^2.\ee 
These additive gradient operators then serve to find directions  in which responses increase, thus enabling a maximal descent algorithm in function space  \cp{mull:10:5}.\vs

\noi{\bf Polynomial Functional Regression.}
Finally, just as the common linear model can be embedded in a more general polynomial version, a polynomial functional model that extends the functional linear model has been developed in \ci{mull:10:3}, with quadratic functional regression as the most prominent social case. With centered predictor processes $X^c$, this model can be written as
\be \label{fqm} E(Y\mid X)=\alpha+\int
\beta(t)X^c(t)dt+\int  \int \gamma(s,t)X^c(s)X^c(t)ds dt,
\ee
and in addition to the parameter function $\beta$ that it shares with the functional linear model it also features a parameter surface $\gamma$. The extension to higher order polynomials is obvious. These models can be equivalently represented as polynomials in the corresponding FPCs. 
A natural question is whether the linear model is sufficient or needs to be extended to a model that includes a quadratic term. A corresponding test was developed by \ci{horv:13}.

\subsection{Time Warping, Dynamics and Manifold Learning for Functional Data}

In addition to amplitude variation, many functional data are best described by assuming that additional time variation is present, i.e, the time axis is distorted by a smooth random process. A classical example are growth data. In human growth, the biological age of different children varies and this variation has a direct bearing on the growth rate that generally follows similar shapes but with subject-specific timing.\vs

\noi {\bf Time Variation and  Curve Registration.} If both amplitude and time  variation are jointly present, they cannot be separately identified, so additional assumptions that break the non-identifiability are crucial if one wishes to identify and separate these two 
 components, which  jointly generate the observed variation in the data. An important consequence of the presence of time warping is that it renders the cross-sectional mean function inefficient and uninterpretable, because if functions have important features such as peaks at different times, ignoring the differences in timing when taking a cross-sectional mean will distort these features. Then the mean curve will not resemble any of the sample curves and is not useful as a 
 representative for the sample of curves \cp{rams:98}.

Early approaches to time-warped functional data included dynamic time warping \cp{sako:78, wang:97} for the registration of speech 
and self-modeling nonlinear regression \cp{lawt:71,knei:88}, where in the simplest case one assumes that the observed random functions can be modeled as shift-scale family of an unknown template function, where shift and scale are subject-specific random variables.
Another traditional method to deal with  time warping in functional data, which is also referred to as the registration or alignment problem, is the landmark method. In this 
approach  special features such as peak locations in functions or derivatives are aligned to their average location and then smooth transformations from the average location to the location of the feature for a specific subject are introduced \cp{knei:92,gass:95}. If well-expressed features are present in all sample curves, the landmark method serves  as a gold standard for curve alignment.  However, landmark alignment requires that all landmarks are present and identifiable in all sample curves. This is often not the case for noisily recorded functional data.  Landmarks may also be genuinely missing in some sample functions due to stochastic variation. 

The mapping of latent bivariate time warping and amplitude processes into random functions can be studied systematically, leading to the definition of the mean curve as the function that corresponds to the bivariate Fr\'echet mean of both time warping and amplitude processes  \cp{mull:04:4} and this  can be exemplified with a simple approach of defining time warping functions by relative area-under-the curve. Recent approaches include alignment of function by means of minimizing a Fisher-Rao metric \cp{wu:14}, alignment of event data by dynamic time warping \cp{mull:14:2}, and time warping in house price boom and bust modeling \cp{mull:14:3}.\vs

\noi {\bf Pairwise Warping.}
As a specific example of how a warping approach can be developed, we discuss a 
 pairwise warping approach that is based on the idea that all relevant information about time warping resides in pairwise comparisons and the resulting pairwise relative time warps  \cp{mull:08:6}. Starting with a sample of $n$ i.i.d.  smooth observed curves $Y_1, Y_2 , . . . , Y_n$ (with suitable modifications  for situations where 
 the curves are not directly observed but only noisy measurements of the curves at a grid of discrete time points are available) we postulate that
 \begin{equation}
Y_i(t)=X_i\{h_i^{-1}(t_j)\}, \,
t \in [0,T],
 \label{eq:yi}\end{equation}
 where the $X_i$ are i.i.d. random functions that represent amplitude variation and the $h_i$ are the realizations of a time warping process $h$ that yields warping functions that represent time variation, are strictly monotone and invertible and satisfy  $h_i(0)=0, \, h_i(T)=T.$ The time warping functions map time onto warped time and since time flows forward only, have to be strictly monotone increasing, A recent approach to warping that allows time to flow backwards with possibly declining warping functions as well has been 
 applied to housing prices where declines correspond to reversing time \cp{mull:14:3}. 
 
To break the non-identifiability, \ci{mull:08:6} (from which the following descriptions are taken) make the assumptions that the overall curve variation is (at least asymptotically) dominated by time 
 variation, i.e., $X_i(t)=\mu(t)+\delta Z_i(t),$ where $\delta$ vanishes for increasing sample size $n$,  the $Z_i$ are realizations of a smooth square integrable process and 
 $E\{h(t)\}=t$, for $t\in [0,1]$. Then warping functions may be represented  in a suitable basis that ensures monotonicity 
 and has associated random coefficients in the expansion, for example monotonically restricted piecewise linear functions. 
 If curve $Y_i$ has the associated time warping function $h_i$ then 
 the  warping function $g_{ik}$ that transforms  the time scale of
curve $Y_i$ towards that of $Y_k$ is  
$g_{ik}(t)=h_i\{h_k^{-1}(t)\}$,
and analogously, the pairwise-warping function of curve $Y_k$ towards
$Y_i$ is $g_{ki}(t)=h_k\{h_i^{-1}(t)\}$.  

Because warping functions are assumed to have average identify, $
E[h_i\{h_k^{-1}(t)\}\bigl|h_k]=h_k^{-1}(t),$ and, as
$g_{ik}(t)=h_i\{h_k^{-1}(t)\}$, we find that
 $   h_k^{-1}(t)=E\{g_{ik}(t)\bigl|h_k\},$
which 
motivates corresponding  estimators by plugging in estimates of the pairwise warping functions. This shows that under certain regularity assumptions the 
relevant warping information is indeed contained in the pairwise time warpings. 

Promising recent extensions of warping approaches aim at formulating joint models for amplitude and time variation or for combinations of regression and time variation  \cp{knei:08,gerv:15, mull:15:1}. Adopting a joint perspective  may lead to better  interpretability in language warping or better performance in functional regression in the presence of warping.\vs

\noi {\bf Functional Manifold Learning.}
 A  comprehensive approach to time warping and other nonlinear features of functional data such as scale or scale-shift families that simultaneously handles amplitude and time warping features is available through manifold learning. A motivation for the use of functional manifold models is  that image data that 
 are  dominated by random domain shifts lie on a manifold \cp{dono:05}. Similar warping models where the warping corresponds to a random time shift have been studied for functional data \cp{silv:95, mull:06:8}. Such data have low-dimensional representations in a transformed space but are infinite-dimensional in the traditional functional basis expansion including the eigenbasis expansion (\ref{KL}).  While these expansions will always converge in $L^2$ under minimal conditions, in these scenarios they lead to an inefficient functional representation in contrast to representations that take advantage of the manifold structure.

When functional data include time warping or otherwise lie on a nonlinear low-dimensional manifold that is situated within the ambient 
 infinite-dimensional functional Hilbert space, desirable low-dimensional representations are possible through manifold learning and the resulting nonlinear representations are particularly useful for
subsequent statistical analysis.  Once a map from an
underlying low-dimensional vector space into functional space has been determined, this gives
the desired 
manifold representation. 
Nonlinear dimension reduction methods, such as locally linear
embedding \cp{rowe:00}, isometric mapping with Isomap  \cp{Tene:00}  and
Laplacian eigenmaps \cp{belk:03} have been successfully applied
to image data and are particularly useful for time-warped functional data, and also for 
samples of random density functions \cp{knei:01, mull:11:4} or other forms of functional data that contain nonlinear structure.
In terms of diagnostics, indicators  for the presence of functional manifolds are plots of FPCs against other FPCs that exhibit 
``horseshoe" or other curved shapes.

Among the various manifold learning methods, Isomap can be easily implemented and has been 
shown to be a useful and versatile method for functional data analysis. Specifically,  a modified Isomap learning algorithm that 
includes  a penalty to the empirical
geodesic distances to correct for noisy data,
and employing local smoothing to map data from the manifold into functional space
has been shown to provide a flexible and broadly applicable approach to low-dimensional manifold modeling of time-warped functional data 
\cp{mull:12:1}. 
This approach targets ``simple'' functional manifolds
$\mani$ in $L^2$ that are ``flat'', i.e.,  isomorphic to a subspace
of Euclidean space, such as a Hilbert space version of the ``Swiss Roll". An essential input for Isomap is the distance between functional data. 
A default distance is the $L^2$ distance, but this distance is not always feasible, for example when the functional data are only sparsely sampled. 
In such cases, the $L^2$ distance needs to be replaced by a distance that adjusts to  sparsity \cp{mull:08:8}.

The manifold $\mani$ is characterized by a  
coordinate map $\varphi: \R^d\rightarrow \mani \subset L^2$, such
that $\varphi$ is bijective, and both $\varphi$, $\varphi^{-1}$ are
continuous and isometric. For a random function $X$ the mean  $\mu$ in the $d$-dimensional representation
space and the manifold mean $\Xmu$ in the functional $L^2$ space are characterized by 
$$\mu=\E\{\varphi^{-1}(X)\}, \quad \Xmu=\varphi^{-1}(\mu).$$ The isometry of the map
$\varphi$ implies that the manifold mean $\Xmu$ is uniquely defined.

In addition to obtaining a mean, a second basic task in FDA is to quantify variation. In analogy to the modes of variation that are available through eigenfunctions and FPCA \cp{cast:86,jone:92},  one can  define {\it manifold modes of variation} 
\begin{equation*} \label{2.1.11}
X^{\mathcal{M}}_{j,\alpha} = \varphi\big(\umani+\alpha\slj
\ejmani\big), \ j=1,\ldots,d, \ \alpha \in \R,
\end{equation*}
where the 
vectors $\ejmani\in \R^d$, $j=1,\ldots,d$, are the eigenvectors
of the covariance matrix of $\varphi^{-1}(X)\in \R^d$, i.e.,
$\cov{(\varphi^{-1}(X))}=\sum_{j=1}^d\lambda_j^{\mathcal{M}}(\ejmani)({\ejmani})^T.
$ Here  $\lambda_1^{\mathcal{M}}\geq \ldots \geq
\lambda_d^{\mathcal{M}}$ are the corresponding eigenvalues and the modes are represented by varying 
the scaling factors $\alpha$. 

Each random function $X \in \mani$ then has a unique representation 
in terms of the  $d-$dimensional vector 
$\bs{\vartheta}=(\vartheta_1,\ldots,\vartheta_d)\in \R^d$,
\begin{equation*} \label{2.1.12}
X = \varphi(\umani+\sum_{j=1}^d \vartheta_j \ejmani), \quad
\vartheta_j = \langle\varphi^{-1}(X)-\umani,\ejmani\rangle, \
 j=1,\ldots,d,
\end{equation*}
where  $\langle \cdot,\cdot\rangle$ is the
inner product in $\R^d$ and $\vartheta_j$ are uncorrelated r.v.s
with mean $0$ and variance $\lambda_j^{\mathcal{M}}$, the
functional manifold components \cp{mull:12:1}. This representation is a genuine dimension reduction of the  
functional data to the finite dimension $d$ while the Karhunen-Lo\`eve representation in case of functional data that are on a nonlinear 
manifold in most cases will require an infinite number of components.\vs

\noi {\bf Learning Dynamics From Functional Data.} Since functional data consist of repeated observation of (usually) time-dynamic processes, they allow to determine the dynamics of the underlying processes. Dynamics are typically assessed with derivatives, and under some regularity conditions derivatives $X'$ of square integrable processes $X$ are also square integrable and from the eigenrepresentation (\ref{KL})  (or representation in another functional basis) one obtains 
\be X^{(\nu)}_i(t)= \mu^{(\nu)}(t) + \sum_{k=1}^{\infty} A_{ik} \phi^{(\nu)}_k(t), \label{KL1} \ee
where $\nu$ is the order of derivative. Derivatives of $\mu$ can be estimated with suitable smoothing methods and those of $\phi$ by partial differentiation of 
covariance surfaces, which is even possible in the case of sparsely sampled data where direct differentiation of trajectories would not be possible  \cp{mull:09:1}. 

For the case where one has differentiable Gaussian processes, since $X$ and $X'$ are jointly Gaussian, it is easy to see that \cp{mull:10:2}
 \be \Xd=\beta(t)\{\Xc\} + Z(t), \,\,  \beta(t) =\frac{\cov\{X^{(1)}(t),X(t)\}}{\var\{X(t)\}}. \la{de}
\ee  This is a linear differential equation with a time-varying function $\beta(t)$ and  a drift process $Z$.
Here $Z$  is a Gaussian process such that 
$Z(t),\, X(t)$ are independent at each $t$. If $Z$ is relatively small, the equation is dominated by the linear part and the function $\beta$. 
Then the behavior of $\beta$ characterizes different dynamics, where one can distinguish  {\it dynamic regression to the mean} for those $t$ where  
$\beta(t) <0$ and {\it explosive behavior} for those $t$ where  $\beta(t)>0$. In the first case, deviations of $X(t)$ from the mean function $\mu(t)$ will diminish, while in the second case they will be increase: An individual with a value $X(t)$ above the mean will tend to move even higher above the mean under the explosive regimen but will move closer to the mean under dynamic regression to the mean. Thus the function $\beta$ that is estimated from the observed functional data embodies the empirical dynamics that can be  learned from the observed sample of Gaussian random trajectories.

A nonlinear version of dynamics learning can be developed for the case of non-Gaussian processes 
\cp{mull:12:6}.  This is of interest whenever linear dynamics is not applicable. Nonlinear dynamics learning is based on the fact that one always  has a 
function $f$ such that 
\begin{eqnarray}\label{model}
E\{X'(t) \mid X(t)\}=f\{t,X(t)\}, \quad   X'(t)=  f\{t,X(t)\}+ Z(t)\ ,
\end{eqnarray}
with $E\{Z(t) \mid X(t)\}=0$ almost surely. Generally the function  $f$  will be unknown. It can be consistently estimated from the observed functional data by 
nonparametrically regressing derivatives $X'$ against levels $X$ and time $t$. This can be implemented with simple smoothing methods. The dynamics of the processes is then jointly determined by the function $f$ and the drift process $Z$. Nonlinear dynamics  learning is of interest to understand the characteristics of 
the underlying stochastic system and can also be used to determine whether individual trajectories are ``on track'', for example in applications to
growth curves.

\section{Outlook and Future Perspectives}

FDA has grown from a methodology with a relatively narrow focus on a sample of fully observed functions to encompass other statistical areas that were considered separate. Its applicability is steadily growing and now includes much of  longitudinal data analysis, providing a rich nonparametric methodology for a field that has been dominated by parametric random effects models for a long time. Of special interest are recent developments in the interface of high-dimensional and functional data. There are various aspects to this interface: Combining functional elements with high-dimensional covariates, such as  predictor times within an interval having an individual predictor effect that goes beyond the functional linear model \cp{knei:11},  or selecting arbitrary subsets of functional principal component scores  in regression models.  

Another interface pf high-dimensional and functional data is the method of Stringing \cp{mull:10:4, mull:11:3}, which 
uses a uni- or multi-dimensional scaling  step to order predictors along locations on an interval or low-dimensional domain and then assigns the value of the respective predictor to the location of the predictor on the interval, for all predictors. The distance of the predictor locations on the interval matches as closely as possible a distance measure between predictors that can be derived from correlations.  Combining locations and predictor values and potentially also adding a smoothing step then converts the high-dimensional data for each subject or item to a random function. These functions can be summarized through their FPC scores, leading to an effective dimension reduction that is not based on sparsity and that works well for strongly correlated predictors. These functions can serve as predictors in the framework of one of the above described functional regression  models.  Thus, FDA approaches can take  advantage of the high-diemensional setting of the data and turns it into an advantage.

  Several open problems were mentioned in Section 2 including the  choice of the number of components $K$ needed for the approximation (\ref{KLK}) of the full  Karhunen-Lo{\'e}ve  expansion (\ref{KL}) and   the choice of the tuning parameters involved in the smoothing steps of estimation.  Another less-developed area in FDA is outlier detections and robust FDA approaches.  In general, approaches for sparse functional data are still lagging behind those for dense functional data.  

Many recent developments in FDA have not been covered in this review. These include functional designs and domain selection problems and also dependent functional data such as functional time series, with many recent interesting developments, e.g. \ci{pana:13}. Another area that has gained recent interest are 
multivariate functional data. Similarly, in some longitudinal studies one observes for each subject repeatedly  observed and therefore dependent functional data rather than scalars. There is also recently rising interest in spatially indexed functional data. 
These problems pose novel challenges for data analysis \cp{horv:12}. 

While this review has focused on concepts and not on applications. As for other growing statistical areas, a driving force of recent 
developments in FDA has been the appearance of new types of data that require adequate methodology for their analysis.  This is leading to ``next generation'' functional data that include
more complex features than the  first generation functional data that have been the emphasis of this review. Examples of recent applications include continuous tracking and monitoring of health and movements, temporal gene  expression trajectories, transcription factor count modeling along the genome, and the analysis of auction data, volatility and other financial data with functional methods. Last but not least, brain imaging data are intrinsically functional data and there is an accelerated interest in the neuroimaging community to analyze neuroimaging data with the FDA approach. A separate entry (by John Aston) in this issue deals specifically with this application area.


 \ed
 
 Modeling the covariate effects on the variance components have also been considered for next generation functional data by \citet{MorrVB03, MorrC06, DiCC09, ReisO10, GrevCC10, ScheSG14, ZhuFK14}  and \citet{YuanGG14}.      
 
 Reiss Ogden (2010) image 

Mention topics not covered here sampling and design of functional data \citet{CardDJ13}, domain selection (Hooker? and Mueller?)  incorporating cov in estimation of  the mean (Carrol, Lin and Naisyin...).

Mention all the books and tools for FDA, plus next generation FD.

http://www.stat.colostate.edu/~piotr/reF.pdf

This book presents recently developed statistical methods and theory required for the
application of the tools of functional data analysis to problems arising in geosciences,
finance, economics and biology. It is concerned with inference based on second order
statistics, especially those related to the functional principal component analysis. While
it covers inference for independent and identically distributed functional data, its
distinguishing feature is an in depth coverage of dependent functional data structures,
including functional time series and spatially indexed functions. Specific inferential
problems studied include two sample inference, change point analysis, tests for
dependence in data and model residuals and functional prediction. All procedures are
described algorithmically, illustrated on simulated and real data sets, and supported by a
complete asymptotic theory.
The book can be read at two levels. Readers interested primarily in methodology will
find detailed descriptions of the methods and examples of their application. Researchers
interested also in mathematical foundations will find carefully developed theory. The
organization of the chapters makes it easy for the reader to choose an appropriate focus.
The book introduces the requisite, and frequently used, Hilbert space formalism in a
systematic manner. This will be useful to graduate or advanced undergraduate students
seeking a self-contained introduction to the subject. Advanced researchers will find novel
asymptotic arguments.

Topics not covered - domain selection (Degras and Cardot), http://projecteuclid.org/euclid.bj/1383661214
depth of functional data S L�pez-Pintado, J Romo - Journal of the American �, 2009
\bigskip

\references

\end{document}